\pdfoutput=1

\documentclass[11pt]{article}

\usepackage[final]{acl}

\usepackage{times}
\usepackage{latexsym}

\usepackage[T1]{fontenc}

\usepackage[utf8]{inputenc}

\usepackage{microtype}

\usepackage{inconsolata}

\usepackage{graphicx}

\usepackage{enumitem}
\usepackage{listings}
\usepackage{amssymb}
\usepackage{wrapfig}
\usepackage{booktabs}
\usepackage{amsmath}
\usepackage{tabularx}
\usepackage{adjustbox}
\usepackage{amsthm}
\newtheorem{definition}{Definition}
\usepackage{caption}
\usepackage[ruled,vlined,noend,linesnumbered]{algorithm2e}

\usepackage{kotex}

\usepackage{multirow}   
\usepackage[table]{xcolor}   
\usepackage{colortbl}        

\newcommand{\squishlist}{
   \begin{list}{$\bullet$}
    { \setlength{\itemsep}{1pt}
      \setlength{\parsep}{0pt}
      \setlength{\topsep}{2pt}
      \setlength{\partopsep}{0pt}
      \setlength{\listparindent}{-2pt}
      \setlength{\itemindent}{-5pt}
      \setlength{\leftmargin}{1.5em}
      \setlength{\labelwidth}{0em}
      \setlength{\labelsep}{0.5em} 
    } 
}

\newcommand{\squishend}{
    \end{list}  }

\usepackage{tabularx, array}
\newcolumntype{L}[1]{>{\raggedright\arraybackslash}p{#1}}
\newcolumntype{R}[1]{>{\raggedleft\arraybackslash}p{#1}}

\usepackage{calc} 

\newlength{\Sep}
\newlength{\ColA}\newlength{\ColB}\newlength{\ColC}\newlength{\ColD}
\newcommand{\CellBox}[2]{%
  \begin{minipage}[t]{#1}\raggedright\setlength{\parskip}{0pt}\strut #2\strut\end{minipage}%
}

\newcommand{\TextRow}[4]{%
  \noindent
  \CellBox{\ColA}{#1}%
  \hspace{\Sep}\texttt{|}\hspace{\Sep}%
  \CellBox{\ColB}{#2}%
  \hspace{\Sep}\texttt{|}\hspace{\Sep}%
  \CellBox{\ColC}{#3}%
  \hspace{\Sep}\texttt{|}\hspace{\Sep}%
  \CellBox{\ColD}{#4}\par
}
\newcommand{\TextHeader}{%
  \noindent\textbf{
    \CellBox{\ColA}{Deployment}%
    \hspace{\Sep}\texttt{|}\hspace{\Sep}%
    \CellBox{\ColB}{Organization}%
    \hspace{\Sep}\texttt{|}\hspace{\Sep}%
    \CellBox{\ColC}{Operation}%
    \hspace{\Sep}\texttt{|}\hspace{\Sep}%
    \CellBox{\ColD}{Personnel}}%
  \par
}
\definecolor{linkpink}{HTML}{E91E63}
\hypersetup{
  colorlinks=true,
  urlcolor=linkpink,
  linkcolor=linkpink,
  citecolor=linkpink
}
\usepackage[most]{tcolorbox} 
\usepackage{fontawesome}     

\newtcbox{\codebadge}{
  on line,
  arc=2.2pt,
  colback=linkpink!8,
  colframe=linkpink,
  colupper=linkpink,
  boxrule=0.4pt,
  left=6pt,right=6pt,top=2pt,bottom=2pt
}

\newcommand{\updated}[1]{{\textcolor{black}{#1}}}

\title{LILaC: Late Interacting in Layered Component Graph for \\ Open-domain Multimodal Multihop Retrieval}

\author{
  Joohyung Yun \\
  POSTECH \\
  Republic of Korea \\
  \texttt{jhyun@dblab.postech.ac.kr} \\\And
  Doyup Lee \\
  DirectorLabs \\
  United States \\
  \texttt{doyup@directorlabs.ai} \\\And
  Wook-Shin Han \thanks{\ \ Corresponding author.} \\
  POSTECH \\
  Republic of Korea \\
  \texttt{wshan@dblab.postech.ac.kr} \\
}

\begin{document}
\maketitle
\begin{abstract}

Multimodal document retrieval aims to retrieve query-relevant components from documents composed of textual, tabular, and visual elements. 
An effective multimodal retriever needs to handle two main challenges:
(1) mitigate the effect of irrelevant contents caused by fixed, single-granular retrieval units, and 
(2) support multihop reasoning by effectively capturing semantic relationships among components within and across documents. 
To address these challenges, we propose \texttt{LILaC}, a multimodal retrieval framework featuring two core innovations. 
First, we introduce a \textit{layered component graph}, explicitly representing multimodal information at two layers---each representing coarse and fine granularity---facilitating efficient yet precise reasoning. 
Second, we develop a \textit{late-interaction-based subgraph retrieval} method, an edge-based approach that initially identifies coarse-grained nodes for efficient candidate generation, then performs fine-grained reasoning via late interaction.
\updated{Extensive experiments demonstrate that \texttt{LILaC} achieves state-of-the-art retrieval performance on all five benchmarks, notably without additional fine-tuning.
We make the artifacts publicly available at \href{https://github.com/joohyung00/lilac}{\textcolor{linkpink}{\texttt{github.com/joohyung00/lilac}}}.
}
\end{abstract}

\section{Introduction}\label{sec:intro}

\begin{figure}[ht]
  \centering
  \includegraphics[width=\linewidth]{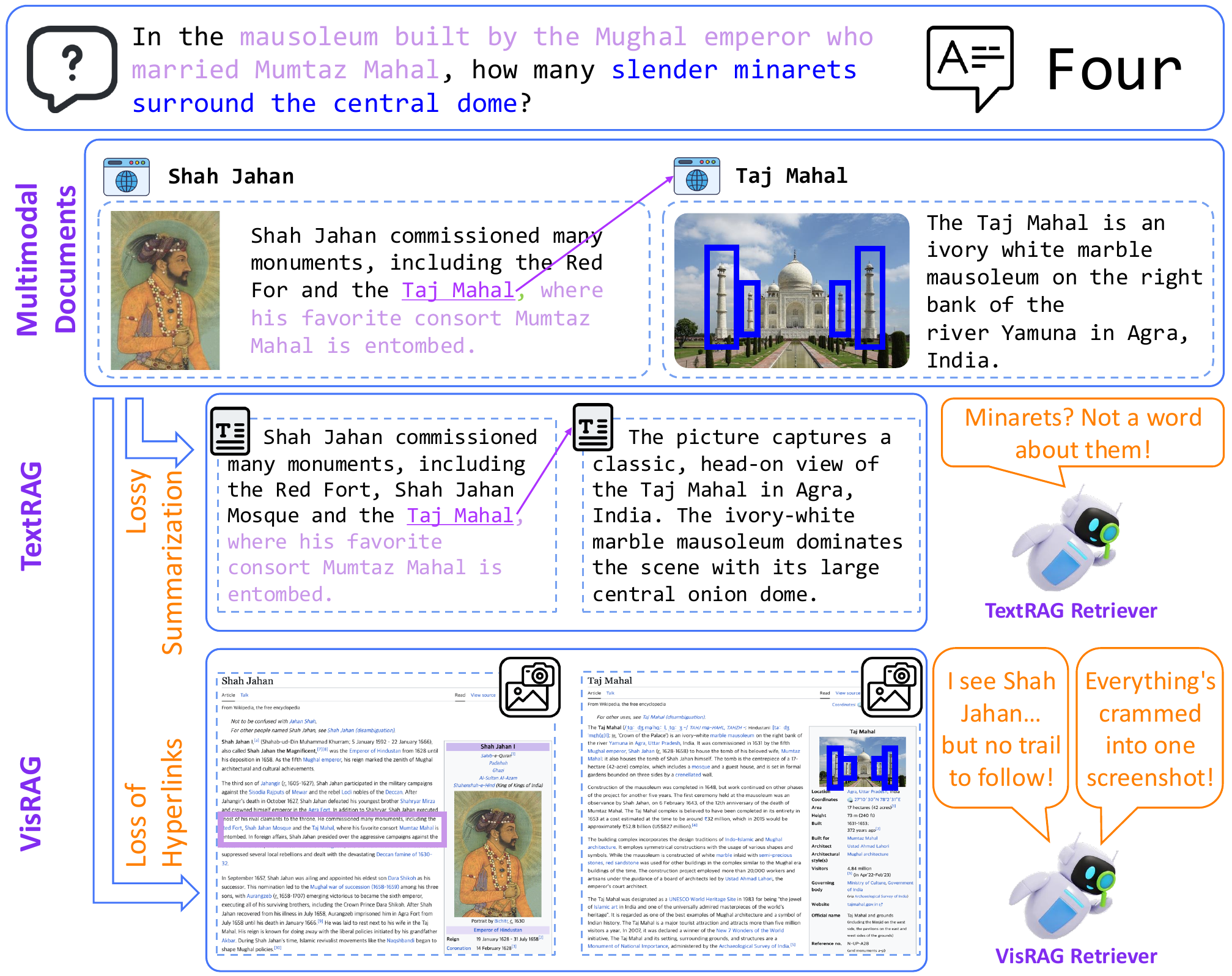}
  \caption{Challenges of TextRAG approaches and VisRAG approaches.
    (a) Incorrect summarization may result in possible information loss in TextRAG.
    (b) Insufficient retrieval granularity in VisRAG.
    (c) Limited multihop reasoning due to loss of links in VisRAG.
  }
  \label{fig:introduction_motivation}
  \vspace{-0.5cm}
\end{figure}

Multimodal retrieval is a rapidly advancing research area, crucial for enhancing modern information retrieval systems~\cite{blip, blip2, clip}. 
Early studies primarily focused on multimodal component retrieval, where components such as text, tables, and images had limited or no explicit relationships~\cite{multimodalqa, webqa}. 
Recently, however, there has been an emerging shift toward open-domain multimodal document retrieval, where closely related components of various modalities are grouped together as a unified document, such as webpages or PDFs~\cite{visrag, m3docrag}.
Such multimodal documents can be viewed as collections of potentially interconnected components (e.g., via hyperlinks as shown with \texttt{Taj Mahal} in Figure~\ref{fig:introduction_motivation}), each belonging to one of multiple modalities, including text, tables, or images.

Recent approaches in multimodal document retrieval have increasingly adopted \textit{VisRAG}-based methodologies, which unify diverse modalities by treating them primarily as visual content, typically represented through screenshots such as a page of a PDF file~\cite{visrag, colpali, m3docrag}. 
By casting multimodal retrieval as essentially an image retrieval problem, these methods leverage advanced vision-based embedding models to preserve multimodal information. 

This paradigm emerged largely as a response to the limitations of earlier \textit{TextRAG}-based approaches, which predominantly relied on textual retrieval by converting visual data into textual summaries~\cite{textrag1, textrag2, textrag3, awakening, skurg, solar, unimmqa}. 
Although effective in leveraging mature text retrieval systems, these methods inherently struggled to represent visual content adequately, resulting in potential information loss and reduction in retrieval effectiveness.
For example, in Figure~\ref{fig:introduction_motivation}, the summary of the \texttt{Taj Mahal}'s image omits the word \texttt{minarets}, which was crucial for answering the query in this context.

Despite their conceptual advances, current multimodal retrieval approaches, including VisRAG, still face two crucial limitations:

\textbf{(1) Insufficient consideration of retrieval granularity.}
Effective retrieval demands explicitly setting an optimal granularity of information representation ~\cite{densexretrieval}. 
Existing VisRAG methods, however, typically adopt a fixed, single-granular approach---generally at the full-page screenshot level---which may include multiple components irrelevant to the query. 
Empirically, we observed that a single screenshot typically comprises an average of three distinct components.
Consequently, the portion of query-relevant information within each screenshot is relatively small, inevitably leading to diminished embedding quality and retrieval effectiveness.
Thus, granularity-aware retrieval remains largely unaddressed within multimodal document retrieval settings.
For example, in Figure~\ref{fig:introduction_motivation}(b), VisRAG struggles because the query-relevant information constitutes only a small portion of the screenshot's content.

\textbf{(2) Limited capability for multihop reasoning.}
Multimodal document retrieval inherently requires reasoning about complex intra- and inter-document relationships among components.
Effective multihop reasoning critically depends on capturing these relationships, as within-document retrieval often necessitates integrating complementary information distributed across multiple modalities to fully represent an entity.
Likewise, inter-document retrieval typically demands traversing semantic connections between related documents. 
Existing VisRAG-based approaches, however, independently embed and retrieve individual screenshots via nearest-neighbor search, thereby overlooking essential interdependencies among components.
Moreover, these methods disregard inherent structural connections within the same document, such as associations among screenshots originating from the same page or hyperlinks explicitly linking different components.
Although some multimodal component retrieval methods have introduced multihop reasoning capabilities~\cite{skurg}, they largely focus on distractor-based closed-domain settings and rely heavily on online reasoning with Large Language Models, significantly limiting their generalization to open-domain multimodal document retrieval scenarios.
For instance, in Figure~\ref{fig:introduction_motivation}(c), VisRAG struggles with multihop reasoning because it does not utilize the structural link from \texttt{Shah Jahan} to \texttt{Taj Mahal}.

To address the challenges, we propose \texttt{LILaC}, an effective multimodal retrieval approach with two novel ideas:

\textbf{(1) Layered component graph construction.}
We first represent the multimodal document corpus as a layered component graph, explicitly designed to capture multimodal information at two distinct granularities. 
This layered graph structure leverages edges to explicitly encode relationships among components within and across documents, thus inherently facilitating effective multihop reasoning.
Additionally, we utilize a layered representation, enhancing retrieval efficiency and effectiveness.
The coarse-grained layer---where textual content is represented as paragraphs, tables as whole entities, and images in their entirety---provides contextual understanding suitable for broad candidate generation.
While in the fine-grained layer---where paragraphs are extracted into  sentences, tables into discrete rows, and images into detected visual objects---enables precise reasoning by decomposing content into finer units. 
Edges in the coarse-grained layer capture semantic associations among components, while edges connecting coarse-grained nodes to their fine-grained subcomponents represent hierarchical containment relationships.

\textbf{(2) Late-interaction-based subgraph retrieval in layered graph.}
At online time, \texttt{LILaC} retrieves a query-relevant subgraph from the layered component graph. 
A key challenge in this step is the combinatorial explosion of candidate subgraphs, resulting from the extensive number of nodes and edges distributed across both granularity layers~\cite{grag}.
To efficiently manage this complexity, we propose a traversal-based subgraph retrieval method on the layered component graph.
Specifically, we first decompose the original query to identify an initial candidate node set at the coarse-grained layer.
We then iteratively perform beam search by traversing connected edges from these initial candidates, dynamically computing relevance scores at each step.
Crucially, since explicitly computing scores for all potential edges would be computationally prohibitive, we leverage the layered structure of both the graph and query decomposition.
In particular, edge scores are computed dynamically via late interaction between the fine-grained subqueries and the fine-grained nodes associated with each candidate edge, effectively utilizing node-level embeddings.

In summary, we make three key contributions: 
(1)  We introduce a layered graph structure capturing multimodal documents at dual granularities, effectively supporting multihop reasoning.
(2) We propose an efficient yet effective subgraph retrieval method leveraging late interaction between decomposed queries and fine-grained components.
\updated{(3) Extensive experiments demonstrate that our approach achieves state-of-the-art retrieval accuracy on all five benchmarks, notably using only pretrained models without additional fine-tuning.}

\section{Preliminary}
\vspace{-2mm}

In this paper, we address multimodal document retrieval, defined as the task of retrieving a ranked list of multimodal components relevant to a given natural language query.
Formally, a retrieval corpus $\mathcal{D}$ comprises a collection of multimodal documents $\{D_1, D_2, \dots, D_{k_{doc}}\}$.
Each multimodal document $D = [C_1, \dots, C_{k_{comp}}]$ is a sequence of multimodal components.
A multimodal component $C$ may belong to one of three distinct modalities
\squishlist
    \item \textit{Paragraph} $P$: a sequence of tokens, forming an unstructured text segment.
    \item \textit{Table} $T$: a structured matrix with rows $T_{i}$ indexed by row number $i$.
    \item \textit{Image} $I$: a tensor $I \in \mathbb{R}^{w \times h \times a}$, with $w$, $h$, and $a$ denote the width, height and the number of channels, respectively.
\squishend
Given a natural language query $Q$, a retrieval corpus $\mathcal{D}$ and a link mapping $\mathcal{L}$, the retrieval task aims to produce a ranked list of components $\mathcal{R} = [C_1, \dots, C_{n_{ret}}]$.
The goal is for the ranked list $\mathcal{R}$ to contain the ground truth set of relevant components ${C_{gt_1}, \dots, C_{gt_r}}$.

The link mapping $\mathcal{L} = \mathcal{C} \to \mathcal{D}$ represents the association or hyperlink relationships between individual components $C$ and their respective multimodal documents $D$, similar to hyperlinks commonly used in webpages and PDF files.

\section{Related Work}

\subsection{Multimodal Document Retrieval}

Early multimodal retrieval methods primarily used a \emph{text-centric} strategy, converting all components—paragraphs, tables, and figures—into plain text, thus losing essential visual cues~\cite{skurg, solar, unimmqa, helios}. 
Later approaches maintained separate embedding spaces for text and images, encoding each modality independently and merging their scores heuristically~\cite{mmragsurvey, beyondtext}. 
However, these methods struggle with reasoning across modalities due to disjoint embeddings.

Recent work pushes modality unification a step further through \textbf{VisRAG} pipelines: documents are rasterized into page- or region-level screenshots, so that paragraphs, tables, and images alike are embedded in a single \emph{visual} space.
\texttt{VisRAG} demonstrates end-to-end vision-based retrieval–augmented generation, while \texttt{ColPali} introduces a late-interaction vision–language model that produces multi-vector page embeddings.
Despite their strengths, VisRAG approaches inherit some limitations.
(i) \textbf{Fixed granularity:} retrieval granularity is fixed as full-page screenshots, which may contain query-irrelevant context.
(ii) \textbf{Limited multihop reasoning:} current pipelines treat each screenshot independently, ignoring the dependencies between components.

\subsection{Granularity of Retrieval}

\begin{figure*}[ht]
  \centering
  \includegraphics[width=\linewidth]{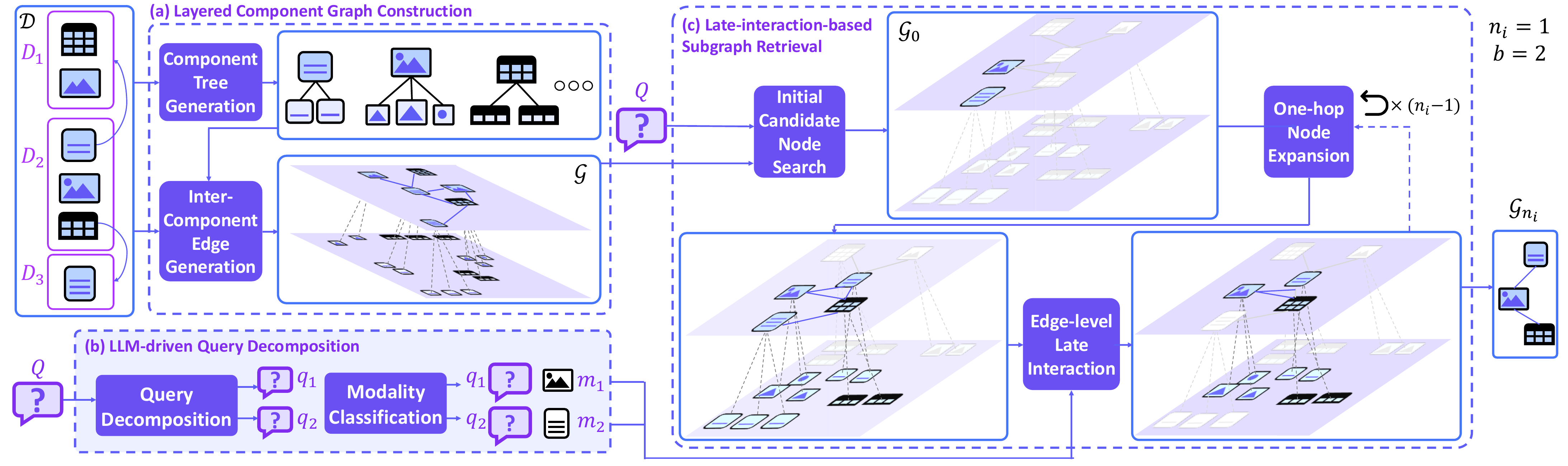}
  \caption{Overview of \texttt{LILaC}.
  (a) A layered component graph is constructed by organizing multimodal documents into coarse- and fine-grained layers. 
  (b) The query is decomposed, followed by modality classification for each subquery.
  (c) \texttt{LILaC} dynamically retrieves a query-relevant subgraph through iterative beam-search traversal. 
  }
  \label{fig:idea_overview}
  \vspace{-0.5cm}
\end{figure*}

Previous studies have explored retrieval granularity across various modalities. 
In text retrieval, \texttt{DenseXRetrieval} demonstrates improved retrieval accuracy using finer sentence- and proposition-level units~\cite{densexretrieval}. 
\texttt{Mix-of-granularity} dynamically selects the optimal granularity tailored to each query~\cite{mixofgranularity}, while \texttt{RAPTOR} starts from sentences and recursively clusters and summarizes them into coarser units~\cite{raptor}. 
For table modality, \textsc{OTT-QA} segments tables into header-plus-row units for targeted row-level retrieval~\cite{ottqa}. 
However, granularity in multimodal document retrieval remains largely unexplored.

\subsection{Multimodal Embedder Models}

Recently, multimodal embedders and their corresponding benchmarks~\cite{vlm2vec, uniir} have emerged as active research areas due to the limitations of traditional uni- or cross-modal embedders in dynamic retrieval scenarios.
Unlike conventional unimodal embedders~\cite{dpr, sentencebert}, multimodal approaches specifically address dynamic settings characterized by retrieval tasks guided by explicit \textit{modality instructions}.
Advanced models such as \texttt{MMEmbed}, \texttt{UniME}, and \texttt{mmE5} leverage sophisticated multimodal language models along with modality-specific fine-tuning, significantly improving retrieval performance under clear modality instructions~\cite{mmembed, unime, mme5}.
However, existing multimodal embedders predominantly focus on training at the component level, leaving the effective use of these models for multimodal document retrieval largely unexplored.
Furthermore, scenarios involving retrieval tasks without explicit instructions or with ambiguous contexts have yet to be thoroughly investigated.

\section{Proposed Method}

We propose \texttt{LILaC}, a novel retrieval algorithm utilizing a layered component graph and traversal method to retrieve a query-relevant subgraph. 
As shown in Figure~\ref{fig:idea_overview}, it consists of two stages: 
(i) \textbf{Layered Graph Construction} organizes multimodal documents into a layered component graph with explicit intra- and inter-document edges. 
(ii) \textbf{Late-interaction-based Subgraph Retrieval} iteratively traverses the layered graph in an edge-wise manner.
To score an edge using node-level embeddings, it uses late interaction between the decomposed subqueries and low-layer subcomponents of an edge.

\subsection{Layered Component Graph Construction}

In the offline phase, \texttt{LILaC} constructs a layered graph structure $\mathcal{G}$, called the \textit{layered component graph}, from the multimodal document set $\mathcal{D}$ and the associated link mapping $\mathcal{L}$.
\updated{
The graph is designed to represent \emph{relationships among components} while also allowing each component to be expressed via \emph{fine-grained constituent elements}.
It comprises two distinct layers explicitly designed to represent semantic relationships among multimodal components, offering two primary advantages.
First, the top layer supports multihop retrieval by explicitly modeling relationships between components and documents, enabling identification of relevant contexts.
Second, the lower layer facilitates precise, fine-grained reasoning by further decomposing components into finer \textit{subcomponents} (defined in Definition~\ref{def:subcomponent}), thus providing detailed context for accurate retrieval.
In addition, the edges explicitly encode two relations among these nodes: 
(i) \emph{hierarchical containment}, which links coarse components to fine-grained subcomponents; 
and (ii) \emph{navigational relations}, which preserve potential cross-component affinity (both intra- and cross-document) without prematurely committing to a specific semantic.}


\begin{definition}[\textbf{Layered Component Graph}]
\label{def:layered_component_graph}
We define a \textit{layered component graph} as $\mathcal{G} = (V, E, \lambda, \tau)$, where $V$ is a set of vertices.
A vertex $v$ belongs to one of the two layers, determined by the layer map \(\lambda : V \to \{0,1\}\), where $0$ and $1$ corresponds to the coarse-grained and fine-grained nodes, respectively.
\vspace{-3mm}
\begin{align*}
    V_0 &= V_{\text{para}}\;\cup\;V_{\text{tbl}}\;\cup\;V_{\text{img}} \\ 
    V_1 &= V_{\text{sent}}\;\cup\;V_{\text{row}}\;\cup\;V_{\text{obj}} 
    \vspace{-3mm}
\end{align*}
We denote each vertex set - \(V_{\text{para}}\): paragraphs, \(V_{\text{tbl}}\): tables, \(V_{\text{img}}\): images, \(V_{\text{sent}}\): sentences, \(V_{\text{row}}\): table rows, \(V_{\text{obj}}\): visual objects detected in images.
The type map
\(\tau : V \to
  \{\texttt{para},\texttt{tbl},\texttt{img},\texttt{sent},\texttt{row},\texttt{obj}\}
\)
refines the vertex set $V$ into the six disjoint categories.
The edge set \(E \subseteq V \times V\) is the union \(E = E_{0} \cup E_\downarrow\) where
\vspace{-3mm}
\begin{align*}
  E_0 &\;=\; \bigl\{(u,v)\in V_0^2\}              \\
  E_\downarrow &\;=\; \bigl\{(u, v)\mid u \in V_0, v \in V_1\}
  \vspace{-3mm}  
\end{align*}
\(E_0\) captures \emph{relationships} between the macro components, while \(E_\downarrow\) captures the containment of a macro component of its subcomponent.
\end{definition}

\begin{definition}[\textbf{Subcomponent}]
\label{def:subcomponent}
Let \(C\) be a multimodal component.  
A \emph{subcomponent} \(c \in \mathcal{S}(C)\) is defined in a modality-specific manner:
\squishlist
    \item \textbf{Paragraph.}  
          For a paragraph \(P = [p_1,\dots,p_{k_{sent}}]\) consisting of sentences, each sentence $p_j$ is a subcomponent.
    \item \textbf{Table.}  
          Let \(T = [T_0;T_1;\dots;T_{k_{row}}]\) where \(T_0\) is the header row.  
          For every data row \(T_i \;(1 \le i \le k_{row})\), the two-row segment $t_i = [\,T_0;\,T_i\,]$ is a subcomponent.
    \item \textbf{Image.}  
          Given an image tensor \(I \in \mathbb{R}^{w \times h \times a}\) and an object detector that
          returns a bounding box \((x_1,y_1,x_2,y_2)\), the corresponding patch
          \vspace{-3mm}
          \[
            i = I[x_1:x_2,\;y_1:y_2,\,:]
            \vspace{-3mm}
          \]
          is a subcomponent.
\squishend
\end{definition}

Layered component graph $\mathcal{G}$ is constructed in two steps.
First, \texttt{LILaC} builds a \textit{component tree} for each component $C$ within $\mathcal{D}$.
A component tree is a two-level tree structure with the root representing the component itself and its children representing the subcomponents, which are extracted differently depending on the modality of the component.
\updated{The roots and leaves of these trees form the nodes of $V_0$ and the nodes of $V_1$, respectively, while the parent–child links correspond to the edges in $E_{\downarrow}$.}
For a paragraph $P$, \texttt{LILaC} utilizes a Sentence-aware Transformer (\texttt{SaT}) model to split it into a set of sentences.
A table $T$ is parsed to generate a set of table segments.
Lastly, a multimodal LLM is used to detect objects within $I$.
\texttt{LILaC} then generates an edge $(C, c) \in E_\downarrow$ for $c \in \mathcal{S}(C)$.

In the next step, \texttt{LILaC} generates the inter-component edges $E_0$ using both inherent structural relationships and hyperlink-based connections. 
For every document $D \in \mathcal{D}$, a clique is formed among its components:
\vspace{-3mm}
\begin{equation}
    E_{intra} = \{(C_i, C_j) | C_i \neq C_j, C_i, C_j \in D\}
    \vspace{-3mm}
\end{equation}
To enable cross-document multihop reasoning, \texttt{LILaC} then follows the link mapping $\mathcal{L}$.
For each pair $(C, D) \in \mathcal{L}$, it connects $C$ to every component in the linked document $\mathcal{D}$.
\vspace{-3mm}
\begin{equation}
    E_{inter} = \{(C, C') | (C, D) \in \mathcal{L}, C' \in D\}
    \vspace{-3mm}
\end{equation}
The inter-component edge set for the top layer is therefore $E_0 = E_{intra} \cup E_{inter}$.
Finally, every node $v \in V$ receives an embedding $\textbf{v} = f(v)$ from a pre-trained multimodal encoder $f$.

\subsection{Late-interaction-based Subgraph Retrieval}

During the online phase, \texttt{LILaC} retrieves a query-relevant subgraph $\mathcal{G}'$ from the layered component graph $\mathcal{G}$ given a query $Q$.  
This retrieval faces two key challenges: 
(1) Direct identification of an optimal subgraph from all possible candidates is computationally infeasible due to a combinatorial explosion~\cite{grag}. 
In particular, the layered component graph contains numerous edges, making explicit embedding of all edges prohibitively expensive in terms of space and computation.
(2) Queries often lack explicit modality instructions, causing ambiguity for multimodal embedders, particularly in complex multihop scenarios~\cite{uniir}.
To address these, we introduce a two-step retrieval strategy: 
(i) \textit{LLM-driven query decomposition}, which explicitly generates modality-specific subqueries, and 
(ii) \textit{Late-interaction-guided graph traversal}, a beam-search traversal method dynamically scoring edges based on fine-grained interactions within the low-level nodes.

\subsubsection{LLM-driven Query Decomposition}

Given a potentially complex query $Q$, \texttt{LILaC} first leverages an LLM to explicitly decompose $Q$ into simpler modality-specific subqueries.
Specifically, we utilize a zero-shot prompting strategy to generate a small set of subqueries:
\vspace{-2mm}
\begin{equation}
\{q_1, \dots, q_{k_{sub}}\} = \text{LLM}(Q;\, prompt_{\mathrm{dec}})
    \vspace{-2mm}
\end{equation}
Each subquery is then classified into a modality label  
$m_j\!\in\!\{\texttt{text},\texttt{table},\texttt{image}\}$ with a second prompt:
\vspace{-2mm}
\begin{equation}
m_j \;=\; \text{LLM}(q_j;\, \textit{prompt}_{\mathrm{mod}})
    \vspace{-2mm}
\end{equation}
Using these labels, we obtain modality-specific embeddings $\mathbf{q}_j = f(q_j;\, m_j)$ for every subquery, while the original query is embedded coarsely as $\mathbf{Q} = f(Q;\, \varepsilon)$ to seed the initial candidate search.  
We denote the set of embedded subqueries as $\textbf{Q}_{\text{sub}} = \{\mathbf q_1,\dots,\mathbf q_{k_{\text{sub}}}\}$.
Full prompt templates appear in \S\ref{sec:prompt_templates}.

\subsubsection{Late-interaction-guided Graph Traversal}
\label{sec:late_interaction}

At inference time, \texttt{LILaC} searches for a subgraph $\mathcal{G}'\!\subseteq\!\mathcal{G}$ that best matches the query.  
\texttt{LILaC} maintains a beam of size $b$ and iteratively identify a candidate subgraph $\mathcal{G}_t=({V}_t,{E}_t, \lambda, \tau)$ consisting of $b$ edges.
Initially, to efficiently narrow the search space from numerous candidate nodes, \texttt{LILaC} identifies a set of top-$b$ top-level nodes $V_0$ most relevant to the query.
\begin{equation}
{V}_{0}
  = \operatorname*{arg\,max}_{C\in V_0}^{b}
    \operatorname{sim}\!\bigl(\mathbf{Q},\mathbf{C}\bigr),
\quad
{E}_{0}= \{\} 
\end{equation}

\texttt{LILaC} then initiates iterative traversal of the graph starting from these candidate nodes. 
In each iteration, \texttt{LILaC} first expands the candidate nodes via one-hop traversal to consider adjacent nodes, dynamically computing query-relevance scores for all edges formed by these expansions. 
Subsequently, only the top-$b$ scored edges are retained for the next iteration forming subgraph, and their constituent nodes become the new set of candidate nodes, forming $\mathcal{G}_i = (V_i, E_i, \lambda, \tau)$. 
After the final iteration $n_i$, \texttt{LILaC} returns the top-$n_{ret}$ nodes from the final subgraph $\mathcal{G}_{n_i}$.

\begin{figure}[ht]
  \centering
  \includegraphics[width=\linewidth]{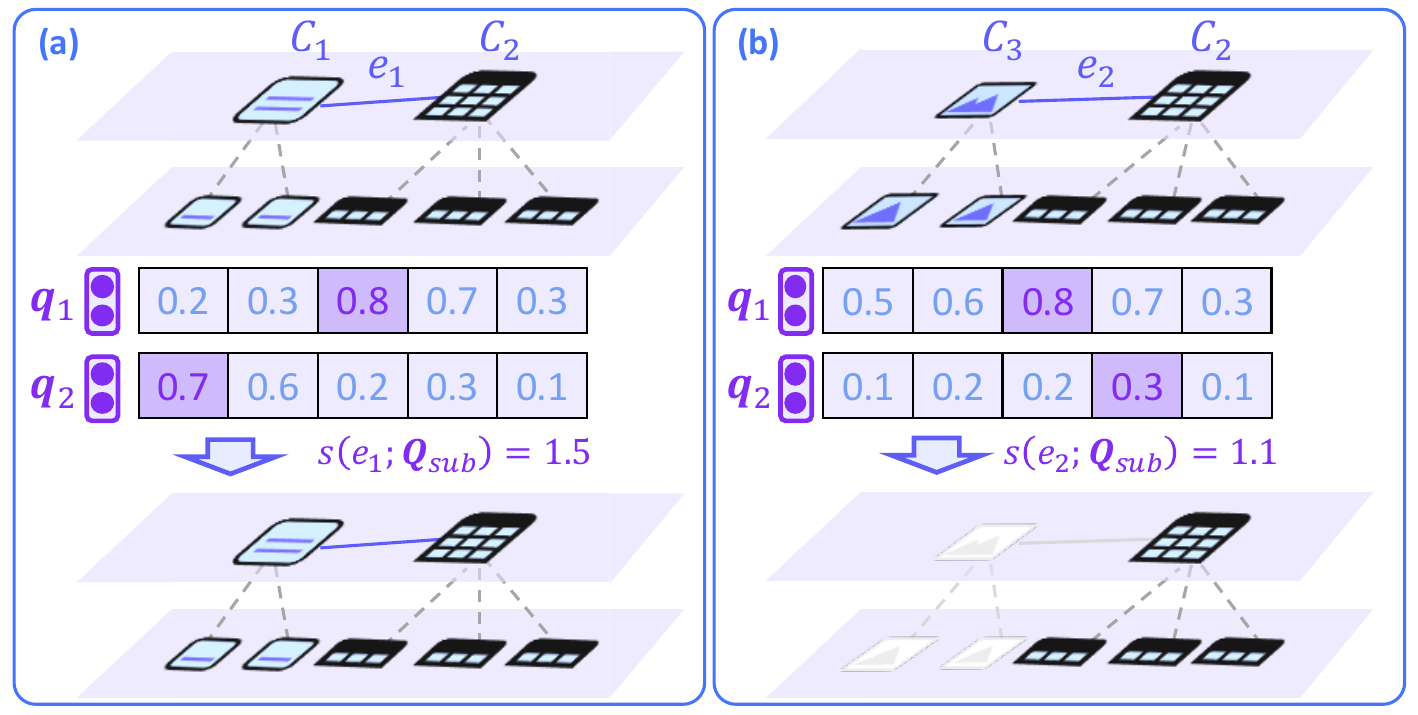}
  \caption{An example case of edge-level late interaction.}
  \label{fig:late_interaction}
\end{figure}

\textbf{Late Interaction Edge Scoring.}
As previously discussed, naively calculating edge scores negatively impacts both effectiveness and efficiency. 
Specifically, this is because 
(1) subqueries, each potentially targeting distinct modalities, must accurately align with the relevant nodes, and 
(2) embedding all edges within the layered graph is inefficient due to their vast number.

To efficiently address these issues, \texttt{LILaC} employs a \emph{late interaction} strategy, scoring each edge on-the-fly with \emph{fine-grained} evidence.
\updated{\texttt{LILaC} extends the standard token-level late interaction to operate at the node-subquery level, by matching decomposed subqueries against the subcomponents contained within an edge.}
Let an edge be $e = (C_\alpha, C_\beta)$ and $\mathcal{S}_e = \mathcal{S}(C_\alpha) \cup \mathcal{S}(C_\beta)$.
\texttt{LILaC} gathers every subcomponent that could provide evidence on either side of the edge in the set $\mathcal{S}_e$.
\vspace{-3mm}
\begin{equation}
s(e;\textbf{Q}_{sub})
    \;=\;
    \sum_{\mathbf q\in\mathbf{Q}_{sub}}
        \max_{c\in\mathcal S_e}
        \operatorname{sim}\bigl(f(c),\mathbf q\bigr).
    \vspace{-3mm}
\label{eq:edge_score}
\end{equation}
The inner \(\max\) selects, for each sub-query \(\mathbf q\), the single most relevant sub-component \(\mathbf c\) incident to the edge, while the outer sum ensures every sub-query contributes exactly once. 
Figure~\ref{fig:late_interaction} shows two example cases of late interaction scoring.
This scoring approach is designed to reflect practical scenarios where each subquery specifically targets fine-grained details located within particular subcomponents.
By aggregating the maximum similarity scores across these detailed elements, rather than relying solely on coarse component embeddings, \texttt{LILaC} effectively prioritizes precise, subcomponent-level matches. 
This strategy enhances retrieval accuracy by focusing directly on relevant information, reducing the noise introduced by broader, less relevant contexts.

We introduce two special cases of edge scoring: 
\textit{(i) Isolated nodes.}  
If a component $C$ has no explicit neighbor, we introduce a dummy edge $(C,\varepsilon)$ so that $C$ can still be considered.
\textit{(ii) One-sided matches.}  
If an edge score $s(e;Q)$ equals the best single-node score of one endpoint, we return only that node to avoid including irrelevant neighbors.
Refer to Figure~\ref{fig:late_interaction} (b) for a specific example.

\section{Experiments} 
\label{sec:exp}
\vspace{-2mm}
\subsection{Experimental Setups}

\begin{table*}[t]
  \centering
  \setlength{\tabcolsep}{4pt}
  \renewcommand{\arraystretch}{1.2}
  \scriptsize

  \begin{tabular}{@{}l|c|cc|cc|cc|cc|cc@{}}
    \toprule
    \multirow{3}{*}{\textbf{Algorithm}}
        & \multirow{3}{*}{\textbf{Embedder Type}}
        & \multicolumn{2}{c|}{\textbf{\texttt{MP-DocVQA}}}
        & \multicolumn{2}{c|}{\textbf{\texttt{SlideVQA}}}
        & \multicolumn{2}{c|}{\textbf{\texttt{InfoVQA}}}
        & \multicolumn{2}{c|}{\textbf{\texttt{MultimodalQA}}}
        & \multicolumn{2}{c@{}}{\textbf{\texttt{MMCoQA}}} \\[-1.5pt]
    \cmidrule(lr){3-12}
        &        & R@3 & MRR@10
                 & R@3 & MRR@10
                 & R@3 & MRR@10
                 & R@3 & MRR@10
                 & R@3 & MRR@10 \\
    \midrule
    \texttt{NV-Embed-v2}  & Text 
                 & 67.85 & 61.91 
                 & 88.49 & 79.55 
                 & 86.21 & 80.86
                 & 60.19 & 67.86 
                 & 46.16 & 41.45 \\
    \midrule
    \texttt{VisRAG-Ret} & \multirow{2}{*}{Image} 
                 & \cellcolor{orange!20}{83.25} & \cellcolor{orange!20}{75.55}
                 & \cellcolor{orange!20}{91.55} & \cellcolor{orange!20}{84.30}
                 & \cellcolor{orange!20}{92.76} & \cellcolor{orange!20}{86.22}
                 & 50.08 & 55.08
                 & 27.63 & 23.75 \\[-1pt]
    \texttt{ColPali}      &  
                 & 80.71 & 74.86 
                 & 89.39 & 81.55 
                 & 88.30 & 82.76
                 & 58.73 & 65.05 
                 & 36.24 & 32.33 \\
    \midrule
    \textbf{\texttt{LILaC} (w/ \texttt{mmE5})} & \multirow{3}{*}{Multimodal}
                 & 61.25 & 55.30 
                 & 77.52 & 68.80 
                 & 75.09 & 69.86
                 & 54.79 & 59.02 
                 & 48.88 & 40.30 \\[-1pt]
    \textbf{\texttt{LILaC} (w/ \texttt{UniME})} &  
                 & 77.83 & 71.42 
                 & 84.35 & 77.93 
                 & 82.28 & 75.27
                 & 58.52 & 61.44 
                 & 49.63 & 42.97 \\[-1pt]
    \textbf{\texttt{LILaC} (w/ \texttt{MM-Embed})} &  
                 & \textbf{83.59} & \textbf{78.75} 
                 & \textbf{92.81} & \textbf{84.43} 
                 & \textbf{93.17} & \textbf{86.83}
                 & \textbf{69.07} & \textbf{75.28} 
                 & \textbf{55.80} & \textbf{50.77} \\
    \bottomrule
  \end{tabular}
  \vspace{-1mm}
    \caption{
    \updated{Retrieval accuracy (Recall@3 (\textit{R@3}) and \textit{MRR@10}) of \texttt{LILaC} and its competitors on five benchmarks.  
  The best score in each column is in \textbf{bold}.  
  The in-domain fine-tuned settings are colored in \colorbox{orange!20}{orange}.}
  }
  \vspace{-3mm}
  \label{tab:retrieval_accuracy}
\end{table*}

\begin{table*}[t]
  \centering
  \setlength{\tabcolsep}{4pt}
  \renewcommand{\arraystretch}{1.2}
  \scriptsize

  \begin{tabular}{@{}l|c|cc|cc|cc|cc|cc@{}}
    \toprule
    \multirow{3}{*}{\textbf{Algorithm}}
        & \multirow{3}{*}{\textbf{MLLM}}
        & \multicolumn{2}{c|}{\textbf{\texttt{MP-DocVQA}}}
        & \multicolumn{2}{c|}{\textbf{\texttt{SlideVQA}}}
        & \multicolumn{2}{c|}{\textbf{\texttt{InfoVQA}}}
        & \multicolumn{2}{c|}{\textbf{\texttt{MultimodalQA}}}
        & \multicolumn{2}{c}{\textbf{\texttt{MMCoQA}}} \\[-1.5pt]
        
    \cmidrule(lr){3-12}
            &    & EM & F1
                 & EM & F1
                 & EM & F1
                 & EM & F1
                 & EM & F1 \\
    \midrule

    \texttt{NV-Embed-v2}  
                 & \texttt{Qwen2.5-VL 7B} 
                 & 56.51 & 63.16 
                 & 53.77 & 64.41 
                 & 60.72 & 63.40 
                 & 37.23 & 43.85 
                 & 28.05 & 34.67 \\[-1pt]
    \midrule
    \texttt{VisRAG-Ret}
                 & \texttt{MiniCPM V2.6}
                 & \cellcolor{orange!20}54.31 & \cellcolor{orange!20}68.86 
                 & \cellcolor{orange!20}43.88 & \cellcolor{orange!20}62.37 
                 & \cellcolor{orange!20}50.83 & \cellcolor{orange!20}57.55
                 & 28.18 & 34.01 
                 & 21.51 & 27.87 \\[-1pt]
    \texttt{VisRAG-Ret}  & \texttt{Qwen2.5-VL 7B} 
                 & \cellcolor{orange!20}65.34 & \cellcolor{orange!20}72.24 
                 & \cellcolor{orange!20}55.03 & \cellcolor{orange!20}66.13 
                 & \cellcolor{orange!20}60.16 & \cellcolor{orange!20}61.93
                 & 22.24 & 25.55 
                 & 16.69 & 20.90 \\[-1pt]
    \texttt{ColPali}      
                 & \texttt{Qwen2.5-VL 7B} 
                 & 64.46 & 71.16 
                 & 53.77 & 64.54 
                 & 58.07 & 60.38 
                 & 23.59 & 27.37 
                 & 18.07 & 22.30 \\[-1pt]
    \midrule
    

    \textbf{\texttt{LILaC} (w/ \texttt{mmE5})}  
                 & \texttt{Qwen2.5-VL 7B} 
                 & 52.96 & 59.53 
                 & 50.89 & 59.07 
                 & 50.12 & 53.12
                 & 40.72 & 47.46 
                 & 33.90 & 40.38 \\[-1pt]

    \textbf{\texttt{LILaC} (w/ \texttt{UniME})} 
                 & \texttt{Qwen2.5-VL 7B} 
                 & 62.43 & 69.40 
                 & 53.05 & 62.89 
                 & 53.39 & 56.86
                 & 43.42 & 49.72 
                 & 33.39 & 40.12 \\[-1pt]
                 
    \textbf{\texttt{LILaC} (w/ \texttt{MM-Embed})} 
                 & \texttt{Qwen2.5-VL 7B} 
                 & \textbf{65.48} & \textbf{72.42}
                 & \textbf{55.57} & \textbf{66.32} 
                 & \textbf{60.91} & \textbf{62.87}
                 & \textbf{44.57} & \textbf{51.97} 
                 & \textbf{36.31} & \textbf{43.22} \\[-1pt]
    \bottomrule
  \end{tabular}
  \vspace{-1mm}
    \caption{
    \updated{End-to-end accuracy (EM and F1) of \texttt{LILaC} and its competitors for the 5 benchmarks.  
  The best score in each column is in \textbf{bold}.
  Generation results corresponding to in-domain fine-tuned settings are colored in \colorbox{orange!20}{orange}.}
  }
  \label{tab:end2end_accuracy}
  \vspace{-3mm}
\end{table*}

\quad 
\textbf{Datasets \& Evaluation Metrics. }
We evaluate on total five benchmarks.
Three are \texttt{VisRAG}-extended open-domain VQA datasets—\texttt{MP-DocVQA}~\cite{mpdocvqa} (industrial documents), \texttt{SlideVQA}~\cite{slidevqa}(presentation slides with multi-hop queries), and \texttt{InfoVQA}~\cite{infovqa} (infographics).
For a realistic webpage retrieval setting, we extend multimodal QA benchmarks (\texttt{MultimodalQA}~\cite{multimodalqa}, \texttt{MMCoQA}~\cite{mmcoqa}) using \texttt{M3DocRAG}'s methodology~\cite{m3docrag}. 
Specifically, we reconstruct webpages from URLs annotated in each component label. 
\texttt{MultimodalQA} comprises 3,235 webpages, each averaging approximately 37 components, corresponding to about 12 PDF pages.
\texttt{MMCoQA} comprises 453 webpages, each averaging approximately 32 components, 11 PDF pages.

Following \texttt{VisRAG}, we evaluate retrieval using Mean Reciprocal Rank at 10 (MRR@10). 
Additionally, we include Recall@3 to assess whether the retrieval component successfully captures relevant information within the top three components, aligning with \texttt{VisRAG}'s experimental design that inputs three components to the generation model.
Further details are explained in \textsection~\ref{sec:implementation_details}.

\textbf{Compared Methods.} 
We employ two SOTA methods of VisRAG approaches - \texttt{VisRAG}, which directly encodes document images via VLMs~\cite{visrag}, and \texttt{ColPali}, which employs late-interaction multi-vector embeddings from document images~\cite{colpali}.
We additionally compare with \texttt{NV-Embed-v2}, a SOTA TextRAG method reported by \texttt{VisRAG}. It utilizes a 7.85B model for embedding textualized components.

\textbf{Applied Multimodal Embedding Models. }
We use three multimodal embedders: \texttt{MM-Embed}~\cite{mmembed}, \texttt{UniME}~\cite{unime} and \texttt{mmE5}~\cite{mme5}.
Details about the embedding models can be further found in \textsection~\ref{sec:appendix_model_details}.

\subsection{Retrieval Accuracy Comparison}
\label{sec:retrieval_accuracy_comparison}

\begin{table*}[t]
  \centering
  \setlength{\tabcolsep}{4pt}
  \renewcommand{\arraystretch}{1.2}
  \scriptsize

  \begin{tabular}{@{}l|l|cc|cc|cc|cc|cc@{}}
    \toprule
    \multirow{3}{*}{\textbf{Embedder Model}}
        & \multirow{3}{*}{\textbf{Variant}}
        & \multicolumn{2}{c|}{\textbf{\texttt{MP-DocVQA}}}
        & \multicolumn{2}{c|}{\textbf{\texttt{SlideVQA}}}
        & \multicolumn{2}{c|}{\textbf{\texttt{InfoVQA}}}
        & \multicolumn{2}{c|}{\textbf{\texttt{MultimodalQA}}}
        & \multicolumn{2}{c }{\textbf{\texttt{MMCoQA}}} \\[-1.5pt]
    \cmidrule(lr){3-12}
        &   & R@3 & MRR@10
            & R@3 & MRR@10
            & R@3 & MRR@10
            & R@3 & MRR@10
            & R@3 & MRR@10 \\
    \midrule

                 

    \multirow{3}{*}{\texttt{mmE5}} 
                & \texttt{LILaC (w/o LCG \& QD)}
                 & 48.90 & 43.97 
                 & 75.91 & 68.13 
                 & 65.60 & 58.55
                 & 42.99 & 46.92 
                 & 41.22 & 34.51 \\[-1pt]
                 
                 & \texttt{LILaC (w/o QD)}
                 & 60.81 & 55.02 
                 & 74.14 & 67.58 
                 & 67.70 & 60.01
                 & 45.15 & 51.12
                 & 44.18 & 36.62
                 \\[-1pt]
                 
                & \textbf{\texttt{LILaC}} 
                 & \textbf{61.25} & \textbf{55.35}
                 & \textbf{76.80} & \textbf{68.99}
                 & \textbf{68.91} & \textbf{61.18}
                 & \textbf{54.78} & \textbf{59.32}
                 & \textbf{48.54} & \textbf{40.22} \\[-1pt]



    \midrule
                 
                 
    \multirow{3}{*}{\texttt{UniME}} 
                 & \texttt{LILaC (w/o LCG \& QD)}
                 & 52.12 & 45.31 
                 & 81.47 & 71.22 
                 & 83.57 & 77.07
                 & 47.68 & 49.06 
                 & 45.78 & 38.41 \\[-1pt]
                 
                 & \texttt{LILaC (w/o QD)}
                 & 77.83 & 71.27 
                 & 83.45 & 75.70 
                 & 85.11 & 78.01
                 & 52.18 & 54.01
                 & 47.11 & 39.85
                 \\[-1pt]
                 
                & \textbf{\texttt{LILaC}} 
                 & \textbf{77.83} & \textbf{71.39}
                 & \textbf{84.35} & \textbf{77.93}
                 & \textbf{85.53} & \textbf{78.81}
                 & \textbf{58.43} & \textbf{61.32}
                 & \textbf{49.45} & \textbf{42.91} \\[-1pt]

    \midrule


                 
                 
    \multirow{3}{*}{\texttt{MM-Embed}} 
                 & \texttt{LILaC (w/o LCG \& QD)}
                 & 75.80 & 69.09 
                 & 92.80 & 82.19 
                 & 90.39 & 83.71
                 & 61.10 & 67.35 
                 & 47.94 & 43.75 \\
                 
                 & \texttt{LILaC (w/o QD)}
                 & 82.23 & 77.75 
                 & 92.27 & 83.20 
                 & 92.17 & 85.53
                 & 63.19 & 69.91
                 & 50.18 & 45.59
                 \\[-1pt]
                 
                 & \textbf{\texttt{LILaC}} 
                 & \textbf{83.59} & \textbf{78.75}
                 & \textbf{92.81} & \textbf{84.43}
                 & \textbf{93.17} & \textbf{86.83}
                 & \textbf{69.07} & \textbf{75.28}
                 & \textbf{55.80} & \textbf{50.77} \\

    \bottomrule
  \end{tabular}
  \vspace{-1mm}
  \caption{
    \updated{Ablation study analyzing retrieval accuracy (Recall@3 and MRR@10) of different \texttt{LILaC} variants.
    Best scores per embedder and dataset are highlighted in bold (\texttt{LCG} = Layered Component Graph, \texttt{QD} = Query Decomposition).
    }
  }
  \label{tab:ablation_study}
  \vspace{-3mm}
\end{table*}

We evaluated retrieval accuracies using Recall@3 (R@3) and MRR@10 across five benchmarks. 
Table~\ref{tab:retrieval_accuracy} summarizes the retrieval performance of \texttt{LILaC} and competing methods. 
\updated{Our results indicate that \texttt{LILaC} achieves state-of-the-art (SOTA) performance on all five benchmarks.
Notably, \texttt{LILaC} outperforms the previous VisRAG SOTA models, \texttt{VisRAG-Ret} and \texttt{ColPali}, by substantial margins of 14.24\% and 11.62\% in R@3, and 15.75\% and 11.74\% in MRR@10, on average, respectively. 
These performance gains are especially prominent on datasets that inherently require fine-grained and multihop reasoning (\texttt{MultimodalQA} and \texttt{MMCoQA}), where the relative improvements in average Recall@3 reached 60.68\% and 31.49\%, and MRR@10 improved by 59.90\% and 45.92\%, respectively.}

Our analysis highlights two key findings:
(i) TextRAG of \texttt{NV-Embed-v2}, consistently shows the lowest retrieval accuracy on visually-dependent VQA datasets that include plots and charts, highlighting inherent limitations in handling visual modalities.
(ii) VisRAG methods notably struggle in webpage retrieval settings (\texttt{MultimodalQA}, \texttt{MMCoQA}), underperforming even when compared to the text-based \texttt{NV-Embed-v2}.
Specifically, the stronger VisRAG model, \texttt{ColPali}, showed accuracy drops against \texttt{NV-Embed-v2}, with reductions of 10.70\% in Recall@3 and 20.96\% in MRR@10. 

\subsection{End-to-end Accuracy Comparison}

We conducted end-to-end question answering (QA) experiments to analyze the impact of retrieval accuracy on downstream QA performance. 
The retrieved results were directly input into a multimodal LLM generator for answer generation, primarily using the \texttt{Qwen2.5-VL 7B} model~\cite{qwen2_5}. 
We limited the number of retrieved units fed into the generator to 3, consistent with the experimental setup of \texttt{VisRAG}.
We additionally provide the results from \texttt{MiniCPM V2.6} for comprehensive comparison, following the original \texttt{VisRAG} pipeline. 
Applied prompts are detailed in \textsection~\ref{sec:prompt_templates}.

\updated{Table~\ref{tab:end2end_accuracy} shows that \texttt{LILaC} achieves SOTA end-to-end accuracy on every benchmark, with average EM and F1 scores of 52.56\% and 59.36\%, respectively. 
This represents substantial improvements of 18.67\% and 19.62\% compared to the previously best-performing VisRAG setup, \texttt{VisRAG} with \texttt{Qwen2.5-VL}, which scored 44.29\% in EM and 49.62\% in F1.}
Overall, the end-to-end QA accuracy trends closely align with retrieval accuracy. 
However, a notable exception arises.
Interestingly, despite \texttt{LILaC (w/ mmE5)} having approximately 8.97\% lower retrieval accuracy (R@3) compared to \texttt{NV-Embed-v2}, its EM score surpasses \texttt{NV-Embed-v2} by 19.71\%. 
This divergence highlights the significant information loss inherent to TextRAG methods, which convert visual content entirely into text, underscoring the importance of preserving visual modalities for effective QA. 

\subsection{Ablation Study}

We performed an ablation study to assess the individual contributions of each key component in our framework to retrieval accuracy.
\updated{
Specifically, we evaluated two variants of \texttt{LILaC} across all three multimodal embedding models.
\texttt{LILaC (w/o QD)} is a variant of \texttt{LILaC} without its query decomposition module, adn thus not incorporating the late interaction score mechanism.
It instead incorporates a two-stage retrieval approach on the layered graph: it first selects the top $b$-nearest neighbor components at the coarse level, and then reranks these components by considering subcomponent-level relevance scores.
\texttt{LILaC (w/o LCG \& QD)} further discards the layered component graph (\texttt{LCG}) structure.
It directly applies a k-nearest neighbor search on individual top-layer components without leveraging finer-grained subcomponents.
}

\updated{
Table~\ref{tab:ablation_study} indicates that incorporating the layered component graph to the simple baseline (\texttt{LILaC (w/o LCG \& QD)}) shows notable average improvements—7.33\% in R@3 and 10.13\% in MRR@10.
Further integrating query decomposition with the late interaction mechanism to \texttt{LILaC (w/o QD)} completes the \texttt{LILaC} algorithm, yielding incremental gains of 3.19\% in R@3 and 4.7\% in MRR@10.
}
While these improvements seem modest, closer inspection reveals significant benefits in datasets requiring complex multihop reasoning, particularly \texttt{MultimodalQA} and \texttt{MMCoQA}. 
Specifically, incorporating \texttt{QD} improves R@3 by an average of 7.40\% and MRR@10 by 10.70\% for these two datasets.
Overall, \texttt{LILaC} is demonstrated to be a generalizable method, evidenced by its consistent performance improvements across all multimodal datasets and embedding models.
This robust trend underscores \texttt{LILaC}'s ability to universally enhance retrieval performance across a variety of multimodal embedding scenarios.

\vspace{-1mm}
\subsection{Algorithm Execution Time}
\vspace{-1mm}

\begin{figure}[t]
\centering
\small
\begin{tabular}{@{}c@{}c@{}}
    { \includegraphics[width=.5\columnwidth]{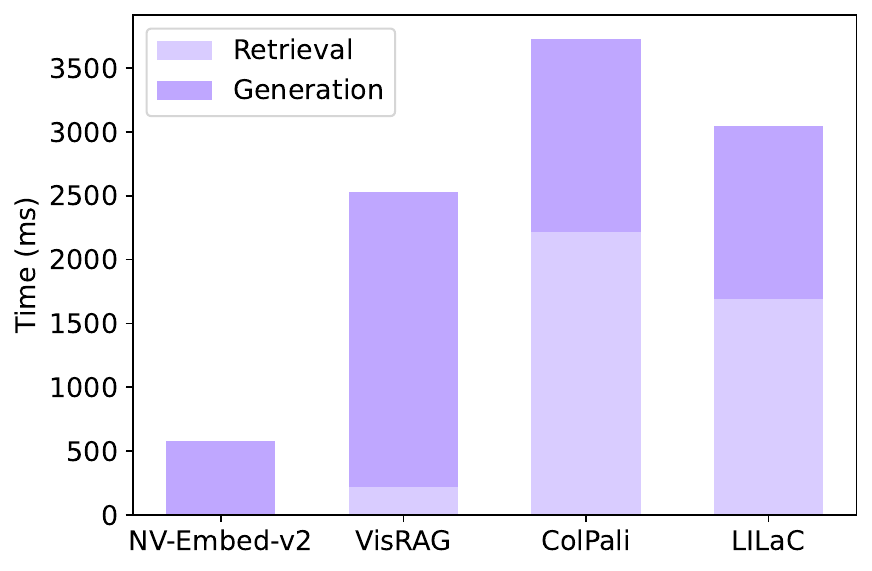}} &
    { \includegraphics[width=.5\columnwidth]{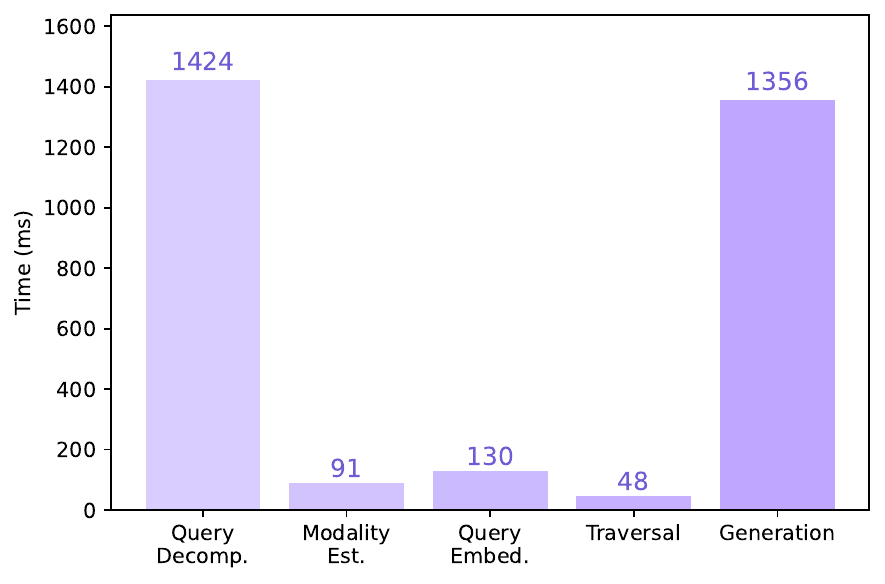}} \\
    \noalign{\vskip -1.3mm}
    (a) Average  runtime &
    (b) Breakdown of \texttt{LILaC} \\
    \end{tabular}
    \vspace{-1mm}
\caption{(a) Comparison of average algorithm execution times across different methods, and (b) detailed runtime breakdown of \texttt{LILaC}.}
\label{fig:algorithm_execution_time}
\end{figure}

Figure~\ref{fig:algorithm_execution_time} (a) shows the average retrieval and generation times for each algorithm.
\texttt{LILaC} is approximately 20.76\% slower than \texttt{VisRAG}, yet 18.24\% faster than \texttt{ColPali}. 
Despite employing a unigranular retrieval approach, \texttt{ColPali}'s runtime remained slower due to its inherent complexity from multi-vector embedding methods.
Notably, both VisRAG methods had longer generation times compared to ours.
\texttt{VisRAG} required 1.70$\times$, and \texttt{ColPali} 1.15$\times$ times our average generation runtime, primarily because their pixel-heavy image inputs increased MLLM inference times.

Figure~\ref{fig:algorithm_execution_time} (b) presents the detailed runtime breakdown for \texttt{LILaC}, showing a total average runtime of 3,047 ms. 
Remarkably, the late-interaction-based subgraph retrieval step accounts for only about 48 ms (approximately 1.5\% of the total runtime). 
The major performance bottleneck lies in the query decomposition phase, averaging 1,423 ms. 
Since this step relies on advanced reasoning with the computationally heavy \texttt{Qwen2.5 72B} model, future improvements in runtime efficiency could be realized by utilizing lighter models, thus balancing speed and retrieval accuracy more effectively.

\vspace{-1mm}
\subsection{Query Decomposition Analysis}
\vspace{-1mm}

\label{sec:exp_query_decomp}
\updated{We evaluate the query decomposition module in isolation and its impact on retrieval accuracy. 
Because benchmarks do not provide gold subqueries, we approximate decomposition quality via the Jaccard similarity between the \emph{predicted} modality set $\widehat{\mathcal{M}}(q)$ (union of modalities assigned to generated subqueries) and the \emph{gold} modality set $\mathcal{M}^\star(q)$ derived from ground-truth components (where $q$ is a query).
Specifically, the score for a query $q$ is $J(q)=\frac{|\widehat{\mathcal{M}}(q)\cap\mathcal{M}^\star(q)|}{|\widehat{\mathcal{M}}(q)\cup\mathcal{M}^\star(q)|}$, and the final accuracy is obtained by averaging over all queries.}

\updated{We compare Qwen2.5 (8B, 72B) and Llama3.1 (7B, 70B) using the prompts of \S\ref{sec:prompt_templates}, keeping all other components fixed. 
In table~\ref{tab:query_decomp_llms}, we report (i) decomposition accuracy, (ii) Recall@3 (R@3), and (iii) decomposition runtime (time (ms)) on \texttt{MultimodalQA} and \texttt{MMCoQA}, as they are the only datasets with $\mathcal{M}^\star(q)$ labeled.
}

\begin{table}[t]
    \centering
    \setlength{\tabcolsep}{4pt}
    \renewcommand{\arraystretch}{1.2}
    \footnotesize
    \begin{tabular}{@{}c|c|c|c|c@{}}
        \toprule
        
        \textbf{LLM} & \textbf{Params} & \textbf{DAcc (\%)} & \textbf{R@3 (\%)} & \textbf{Time (ms)} \\
        
        \midrule
        
        \multirow{2}{*}{Qwen2.5}    & 8B & 63.29 & 59.01 & 258 \\
        \cmidrule(lr){2-5}
        
                                    & 72B & 72.23 & 62.43 & 1849 \\
        \cmidrule(lr){1-5}
        
        \multirow{2}{*}{Llama3.1}   & 7B & 58.11 & 57.44 & 336 \\
        \cmidrule(lr){2-5}
        
                                    & 70B & 66.34 & 61.24 & 1731 \\
                                    
        \bottomrule
    \end{tabular}
    \vspace{-1mm}
    \caption{Query decomposition analysis result on \texttt{MultimodalQA} and \texttt{MMCoQA} datasets.}
    \label{tab:query_decomp_llms}
\end{table}

\updated{
We notice that the LLM-driven decomposition attains reasonable accuracy of 72.23\%, and also that larger models improve both decomposition and retrieval at higher latency. 
Notably, decomposition accuracy and Recall@3 are strongly correlated across model variants (Pearson $\rho{=}0.954$), underscoring that better query decomposition directly benefits retrieval.
}

\vspace{-1mm}
\subsection{Additional Experiments}
\vspace{-1mm}

\updated{
Additional experiments were conducted, but are detailed in the appendix due to space limitations.
These include (i) parameter sensitivity (\textsection~\ref{sec:parameter_sensitivity}), 
(ii) analysis of offline layered component graph construction runtime (\textsection~\ref{sec:lcg_construction_overhead}), and 
(iii) a comparison of retrieval accuracy across different embedder models (\textsection~\ref{sec:embedder_comparison}).
}

\vspace{-1mm}
\section{Conclusion}
\vspace{-1mm}

We presented \texttt{LILaC}, a multimodal retrieval framework designed to address the limitations of existing methods by incorporating layered component graph and late-interaction-based subgraph retrieval.
Our layered graph construction explicitly captures semantic relationships among multimodal components, facilitating effective multihop reasoning. 
The late-interaction retrieval method dynamically evaluates fine-grained component relevance, significantly enhancing retrieval accuracy, yet efficient.
\updated{\texttt{LILaC}'s usage of pretrained multimodal encoders allows it to inherit the improvements from newer off-the-shelf embeddings.
Extensive experiments confirm that \texttt{LILaC} consistently outperforms state-of-the-art approaches across all five benchmarks, also demonstrating its broad applicability and effectiveness in open-domain multimodal retrieval.}

\section{Limitations}

Our current approach focuses on effectively harmonizing pre-trained multimodal models to achieve enhanced retrieval performance without additional fine-tuning. 
Consequently, the accuracy of our retrieval method significantly depends on the quality of subcomponent extraction. 
Also, although our retrieval accuracy surpasses existing methods, there remains substantial room for improvement in end-to-end generation tasks. 

\section*{Acknowledgements}

\updated{
This work was partly supported by the National Research Foundation of Korea(NRF) grant funded by the Korea government(MSIT) (RS-2025-00517736, 50\%), Institute of Information \& communications Technology Planning \& Evaluation (IITP) grant funded by the Korea government(MSIT) (No. RS-2024-00509258, Global AI Frontier Lab, 30\%) (No. RS-2018-II181398, Development of a Conversational, Self-tuning DBMS, 10\%) (No.  RS-2024-00454666, Developing a Vector DB for Long-Term Memory Storage of Hyperscale AI Models, 5\%), and Basic Science Research Program through the National Research Foundation of Korea Ministry of Education(No. RS-2024-00415602, 5\%).
}



\bibliography{custom}

\newpage

\appendix

\section*{Appendix}
\setcounter{section}{0}
\renewcommand\thesection{\Alph{section}}

\section{Software and Data Licenses}

The licenses for the software and datasets used in this paper as follows:

\squishlist
    \item \texttt{VisRAG-Ret}: Apache-2.0
    \item \texttt{ColPali}: PaliGemma License, MIT License
    \item \texttt{MiniCPM-v2.6}: Apache-2.0
    \item \texttt{Qwen2.5-VL 7B}: Apache-2.0
    \item \texttt{Qwen2.5 72B}: Qwen
    \item \texttt{MM-Embed}: CC-BY-NC-4.0
    \item \texttt{NV-Embed-v2}: CC-BY-NC-4.0
    \item \texttt{UniME}: MIT License
    \item \texttt{mmE5}: MIT License
\squishend

All software and datasets were used strictly for research purposes and were not utilized in any non-research contexts, particularly for 
commercial applications.

\section{AI Assistants}

We implemented our code efficiently using ChatGPT-o3~\cite{jaech2024openai}, enabling rapid debugging and effective error resolution. 
Additionally, we revised our paper using ChatGPT-4.5, which helped us enhance sentence clarity and readability through iterative rephrasing.

\section{Reproducibility Statement}

\texttt{VisRAG-Ret} was reproduced using the official code available at \href{https://github.com/OpenBMB/VisRAG}{\textcolor{linkpink}{\texttt{VisRAG official github}}}.

\texttt{ColPali} and \texttt{NV-Embed-v2} were implemented applying their official model cards introduced in  \href{https://huggingface.co/vidore/colpali}{\textcolor{linkpink}{\texttt{ColPali huggingface}}} and \href{https://huggingface.co/nvidia/NV-Embed-v2}{\textcolor{linkpink}{\texttt{NV-Embed-v2 huggingface}}}, respectively.
The source code, data, and other artifacts for \texttt{LILaC} have been made available at \href{https://github.com/joohyung00/lilac}{\textcolor{linkpink}{\texttt{our github repository}}}.

\section{Model Details}
\label{sec:appendix_model_details}

\textbf{(Multimodal) Large language models:}
\squishlist
    \item \texttt{Qwen2.5 72B}: 72B parameters
    \item \texttt{Qwen2.5-VL 7B}: 7B parameters
    \item \texttt{MiniCPM-v2.6}: 8.1B parameters
\squishend

\noindent \textbf{Text embedders}
\squishlist
    \item \texttt{NV-Embed-v2}: 7.85B parameters
\squishend

\noindent \textbf{Cross-modal embedders:}
\squishlist
    \item \texttt{ColPali}: 3B parameters
    \item \texttt{VisRAG-Ret}: 3.43B parameters
\squishend

\noindent \textbf{Multimodal embedders:}
\squishlist
    \item \texttt{MM-Embed}: 8.18B parameters
    \item \texttt{UniME}: 7.57B parameters
    \item \texttt{mmE5}: 10.6B parameters 
\squishend
\texttt{MM-Embed} is fine-tuned via modality-aware hard negative mining~\cite{mmembed}.
\texttt{UniME} is enhanced with textual discriminative knowledge distillation and instruction-tuned hard negatives~\cite{unime}.
\texttt{mmE5} leverages  synthetic multilingual data for robust cross-modal alignment~\cite{mme5}.

\section{Experiment Supplementaries}

\subsection{Hardware and Software Settings}
All our experiments were conducted on a system with an Intel Xeon Gold 6230 GPU @ 2.10GHz, 1.5TB of RAM, and four NVIDIA RTX A6000 GPUs.

\subsection{Implementation Details}
\label{sec:implementation_details}

We set the default hyperparameters for all experiments as beam width $b$ = 30 and number of iterations $n_i$ = 1.
Additionally, for the ablation study that exclusively uses the layered graph structure without late interaction, we also maintained an identical beam width ($b$ = 30) to ensure a fair comparison.

All experiments were conducted with `temperature = 0' and `do\_sample = False'. 
To further ensure fair comparison, we aligned the ratio of components between the VisRAG methods and our approach to approximately 1:3, as justified by the empirical observation that a typical screenshot in our datasets encompasses roughly three distinct multimodal components. 
Specifically, the \texttt{MultimodalQA} dataset contains 39,093 screenshots and 122,521 components, and the \texttt{MMCoQA} dataset comprises 5,175 screenshots and 14,493 components, both yielding a component-to-screenshot ratio close to 3:1.

\subsection{Benchmark Details}

\quad \texttt{\textbf{MP-DocVQA}}: It is a multimodal visual question answering benchmark designed for industrial documents. 
It includes challenging questions that require extracting and reasoning over textual and visual information such as tables, figures, and charts found in documents. 
The development set contains 591 questions sourced from a corpus of 741 multimodal document pages.

\texttt{\textbf{SlideVQA}}: It focuses on extracting information from presentation slides and often requires multihop reasoning across multiple slides. 
It emphasizes the capability to handle diverse layouts and structured textual information commonly found in presentations. 
The \texttt{SlideVQA} development set comprises 556 questions, with the corpus containing 1,284 slide pages.

\texttt{\textbf{InfoVQA}}: It targets visual question answering on infographics, which blend images, charts, and textual descriptions. 
This dataset presents complex multimodal reasoning tasks where models must interpret visual elements combined with succinct textual explanations. 
Its development set includes 718 questions drawn from a corpus of 459 infographic pages.

\texttt{\textbf{MultimodalQA}}: It refers to the extended version of \texttt{MultimodalQA}, with its extension methodology introduced in M3DocRAG~\cite{m3docrag}.
The dataset covers a wide variety of document types, including texts, images, and tables, requiring complex multihop reasoning across multiple documents. 
Its evaluation set comprises 2,441 questions from over 3,368 PDF documents totaling approximately 41,005 pages.

\texttt{\textbf{MMCoQA}}: It is a conversational multimodal question-answering dataset aimed at testing a system’s ability to handle multimodal information across multiple turns in a conversational context. 
This dataset is also an extension of the \texttt{MMCoQA} dataset, which originally operates in a distractor setting.
It involves coherent, multi-turn question sequences requiring integration of information from text, images, and tables. 
The dataset includes 5,753 questions organized into 1,179 conversational dialogues. 
Its corpus consists of 218,285 textual passages, 10,042 tables, and 57,058 images.

\begin{table*}[t]
    \centering
    \small
    \begin{tabular}{l|c|c|c|c|c}
        \toprule
        \textbf{Step} & \textbf{20\%} & \textbf{40\%} & \textbf{60\%} & \textbf{80\%} & \textbf{100\%} \\
        \midrule
        Node Generation       & 2m 8s   & 3m 53s  & 6m 20s  & 8m 29s  & 10m 20s \\
        Edge Generation       & 38s     & 1m 6s   & 1m 46s  & 2m 18s  & 2m 54s  \\
        Embedding Generation  & 24m 27s & 47m 44s & 1h 15m 31s & 1h 40m 42s & 2h 2m 43s \\
        \midrule
        \textbf{Total}        & \textbf{27m 13s} & \textbf{52m 43s} & \textbf{1h 23m 37s} & \textbf{1h 51m 29s} & \textbf{2h 15m 57s} \\
        \bottomrule
    \end{tabular}
    \vspace{-1mm}
    \caption{
    \updated{Average offline construction time by corpus fraction. }
    }
    \label{tab:lcg_offline_runtime}
    \vspace{-2mm}
\end{table*}

\begin{table*}[t]
    \centering
    \small
    \begin{tabular}{l|c|c|c}
        \toprule 
        \textbf{Model} & \textbf{Params} & \textbf{Recall@3 (\%)} & \textbf{MRR@10 (\%)} \\
        \midrule
        \texttt{MM-Embed}
            & 8B  & 78.89 & 75.18 \\   \cmidrule(lr){1-4}        
        \texttt{UniME (LLaVA-OneVision)}
            & 7B  & 69.83 & 65.30 \\ \cmidrule(lr){1-4} 
        \texttt{mmE5-mllama (instruct)}
            & 11B & 62.54 & 57.83 \\ \cmidrule(lr){1-4} 
        \texttt{QQMM-embed}
            & 8B  & 66.09 & 62.00 \\ \cmidrule(lr){1-4} 
        \multirow{3}{*}{\texttt{LLaVE}}
            & 0.5B & 56.59 & 51.76 \\ \cmidrule(lr){2-4} 
            & 2B   & 62.01 & 57.09 \\ \cmidrule(lr){2-4} 
            & 7B   & 67.14 & 62.13 \\ \cmidrule(lr){1-4} 
        \multirow{2}{*}{\texttt{VLM2Vec (Qwen2-VL)}}
            & 2B   & 47.57 & 42.58 \\ \cmidrule(lr){2-4} 
            & 7B   & 53.24 & 47.68 \\
        \bottomrule
    \end{tabular}
    \vspace{-1mm}
    \caption{
    \updated{Retrieval accuracy compared with different pretrained embedders.}
    }
    \label{tab:embedder_bias}
    \vspace{-2mm}
\end{table*}


\subsection{Parameter Sensitivity}
\label{sec:parameter_sensitivity}

\begin{figure}[t]
\centering
\small
\begin{tabular}
{@{}c@{}c@{}}
    { \includegraphics[width=.545\columnwidth]{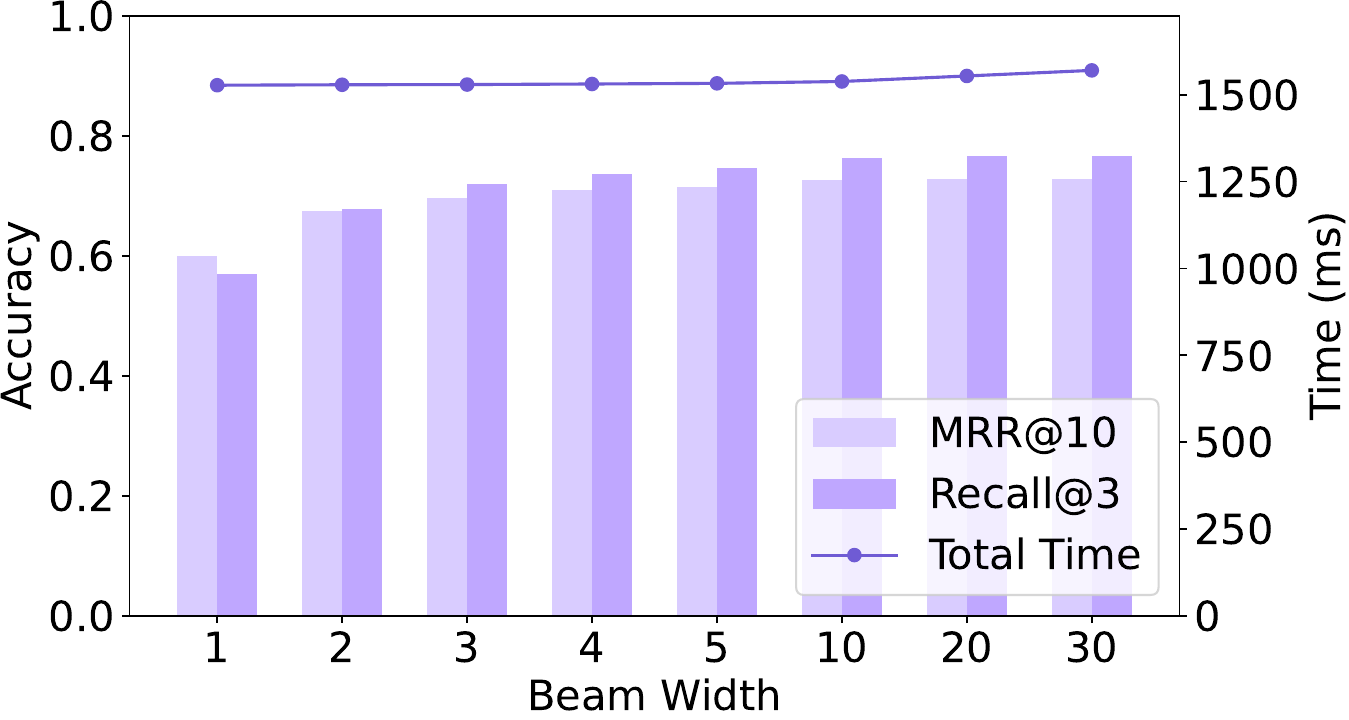}} &
    { \includegraphics[width=.415\columnwidth]{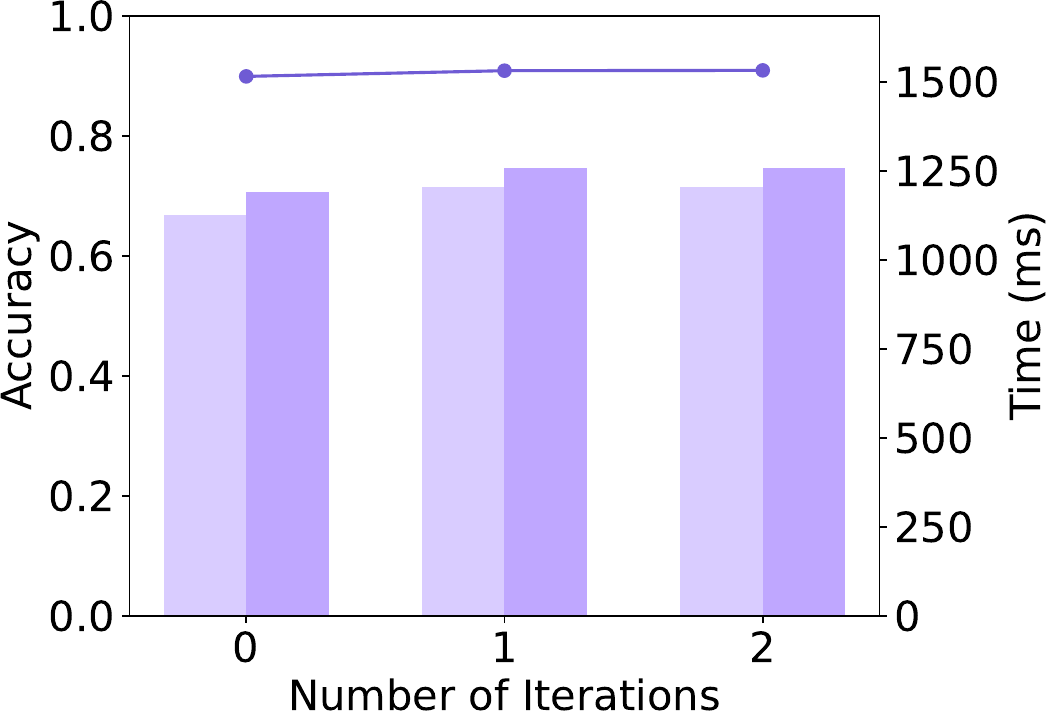}   } \\
    (a) Beam width &
    (b) Number of \\     
                    & 
        iterations \\
\end{tabular}
    \vspace{-3mm}
    \caption{Change in retrieval accuracy with varying parameter values.}
    \label{fig:parameter_sensitivity}
    \vspace{-4mm}
\end{figure}

\begin{figure*}[t]
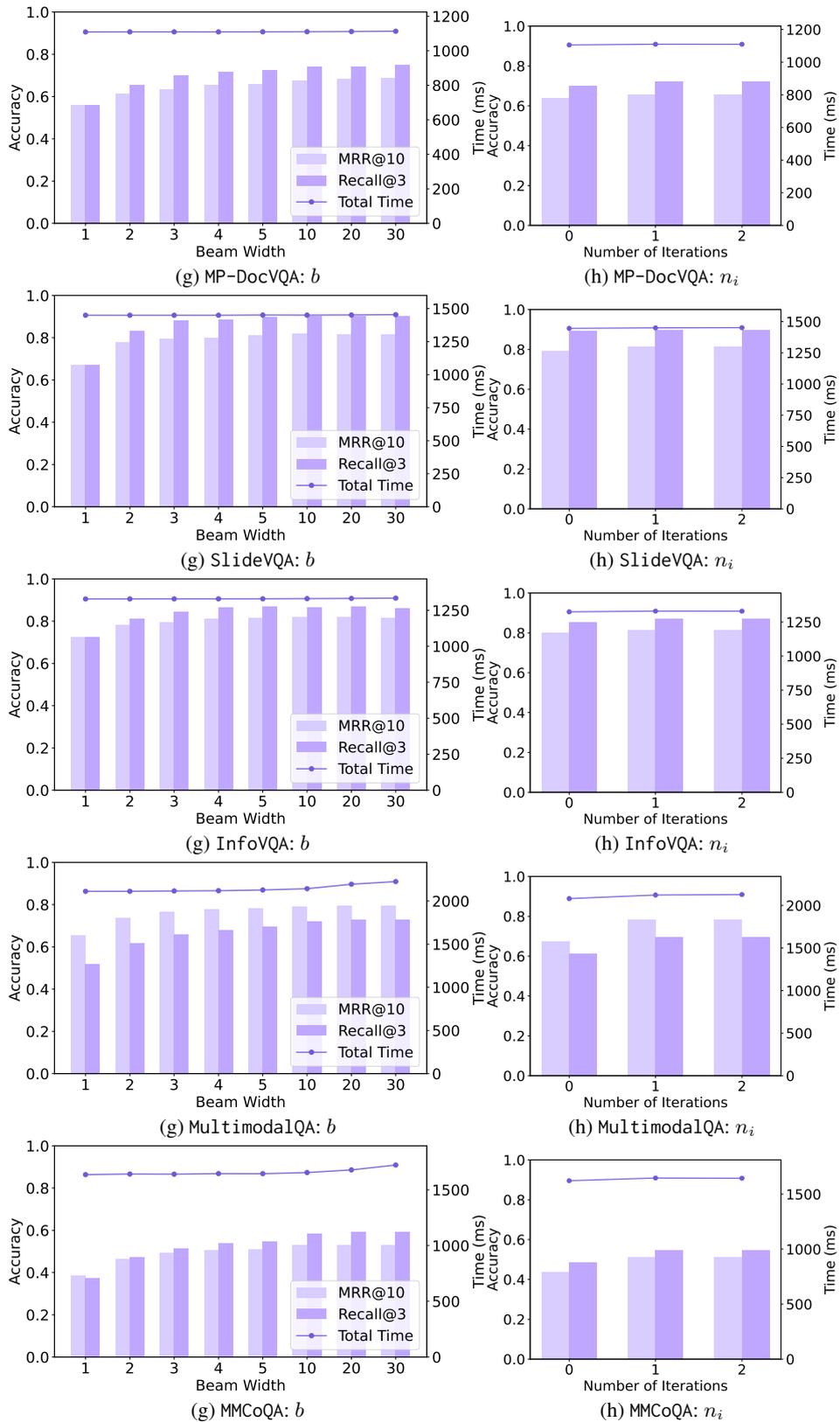

  \centering
  \small
  \newcommand{\imgwa}{0.45\linewidth}   
  \newcommand{\imgwb}{0.33\linewidth}   
  \newcommand{\beam}[1]{\includegraphics[width=\imgwa]{%
      figures/plot_beamwidth/#1.pdf}}
  \newcommand{\iter}[1]{\includegraphics[width=\imgwb]{%
      figures/plot_iterations/#1.pdf}}
  \setlength{\tabcolsep}{3pt}
  \begin{tabular}{@{}c@{}c@{}}

    \beam{MP-DocVQA} & \iter{MP-DocVQA} \\
    (g) \texttt{MP-DocVQA}: $b$ & (h) \texttt{MP-DocVQA}: $n_i$ \\

    \beam{SlideVQA} & \iter{SlideVQA} \\
    (g) \texttt{SlideVQA}: $b$ & (h) \texttt{SlideVQA}: $n_i$ \\
  
    \beam{InfoVQA} & \iter{InfoVQA} \\
    (g) \texttt{InfoVQA}: $b$ & (h) \texttt{InfoVQA}: $n_i$ \\

    \beam{MMWebQA} & \iter{MMWebQA} \\
    (g) \texttt{MultimodalQA}: $b$ & (h) \texttt{MultimodalQA}: $n_i$ \\

    \beam{MMCoQA} & \iter{MMCoQA} \\
    (g) \texttt{MMCoQA}: $b$ & (h) \texttt{MMCoQA}: $n_i$ \\

  \end{tabular}
  \vspace{-3mm}
  \caption{Parameter-sensitivity analysis for each dataset: effect of
           beam width $b$ (left) and number of iterations $n_i$ (right).}
  \label{fig:appendix_param_sensitivity}
\end{figure*}

\begin{figure*}[t]
  \centering
  \small
  \newcommand{\imgw}{0.37\linewidth}     
  \newcommand{\runtime}[1]{\includegraphics[width=\imgw]{figures/plot_runtime/#1.pdf}}
  \newcommand{\breakdown}[1]{\includegraphics[width=\imgw]{figures/plot_breakdown/#1.pdf}}
  \setlength{\tabcolsep}{3pt}            
  \begin{tabular}{@{}c@{}c@{}}
    \runtime{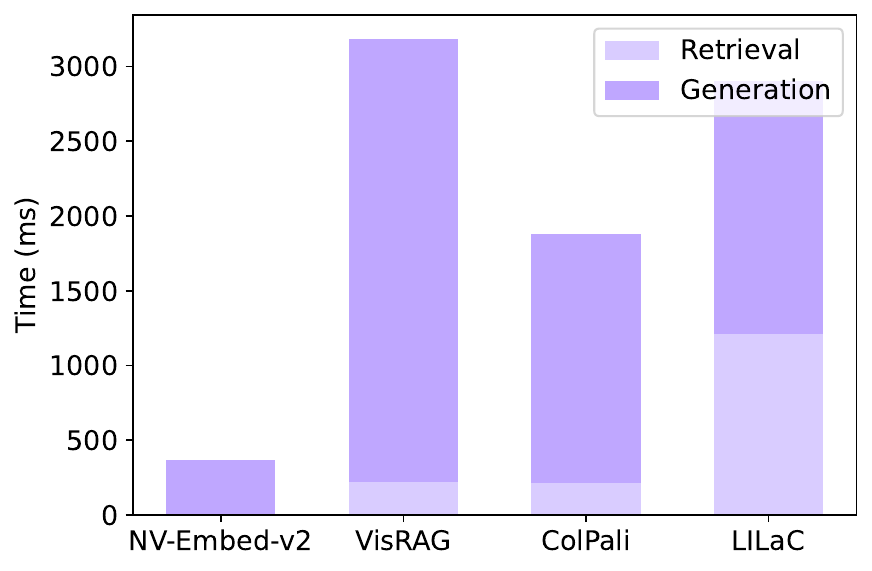} & \breakdown{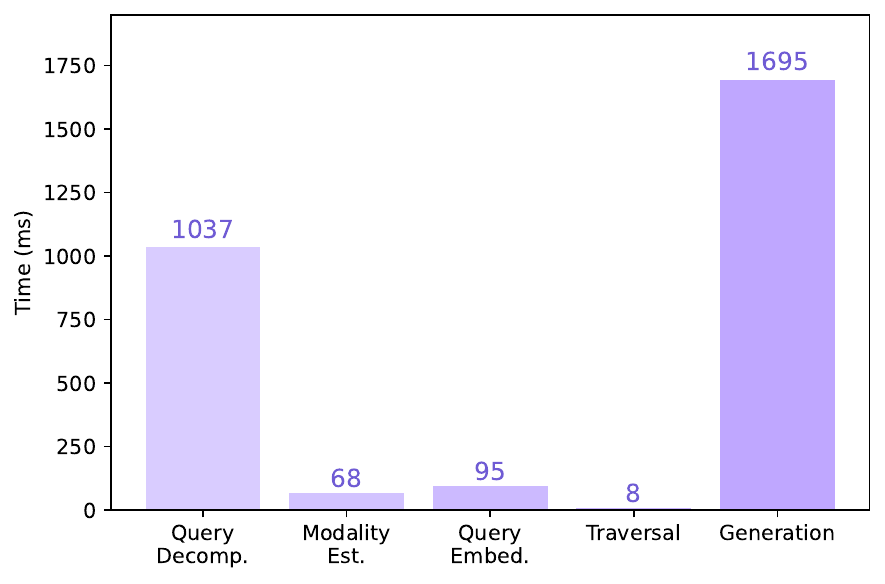} \\
    (a) \texttt{MP-DocVQA} runtime comparison & 
    (b) \texttt{LILaC}'s runtime breakdown on \texttt{MP-DocVQA} \\

    \runtime{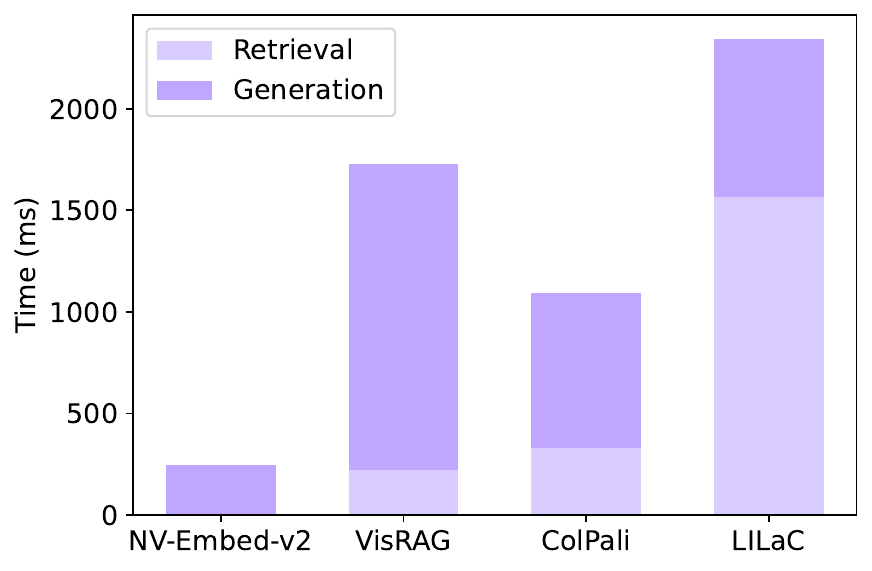} & \breakdown{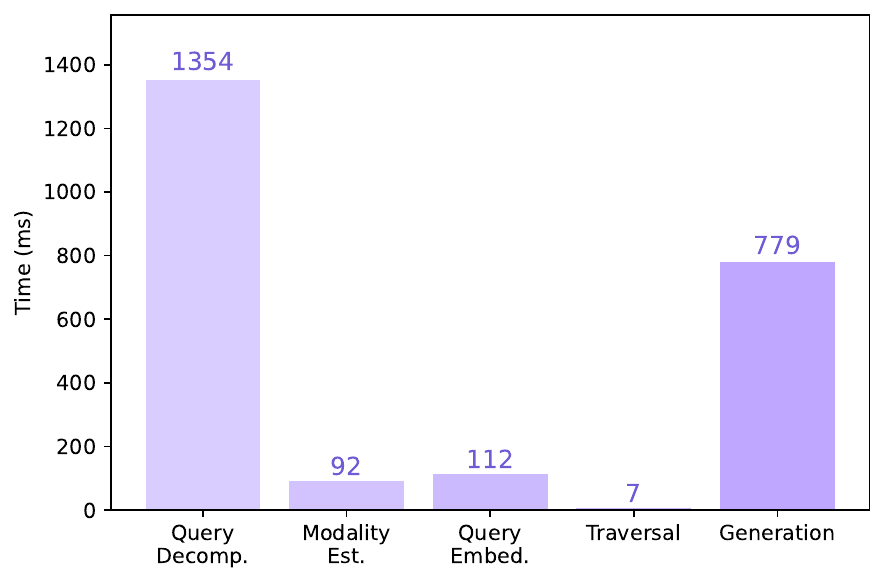} \\
    (c) \texttt{SlideVQA} runtime comparison &
    (d) \texttt{LILaC}'s runtime breakdown on \texttt{SlideVQA} \\
    
    \runtime{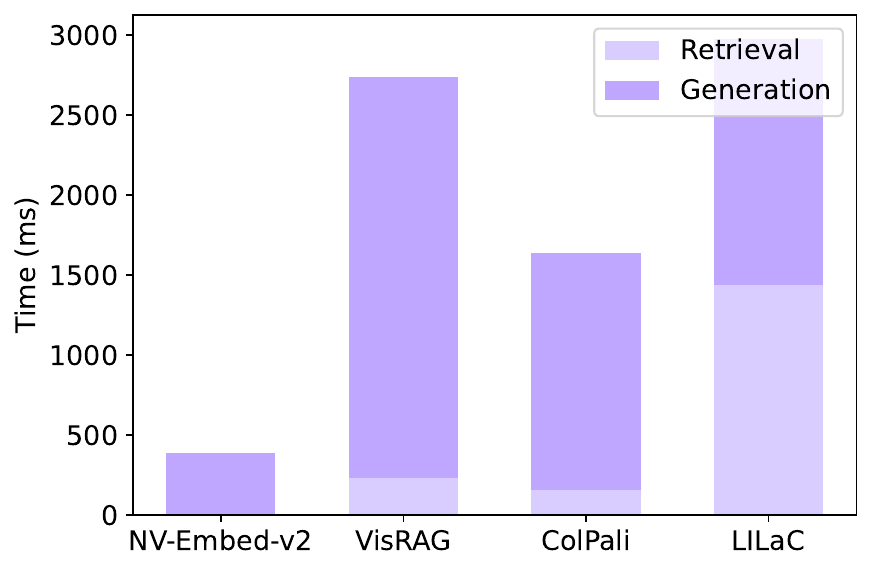}  & \breakdown{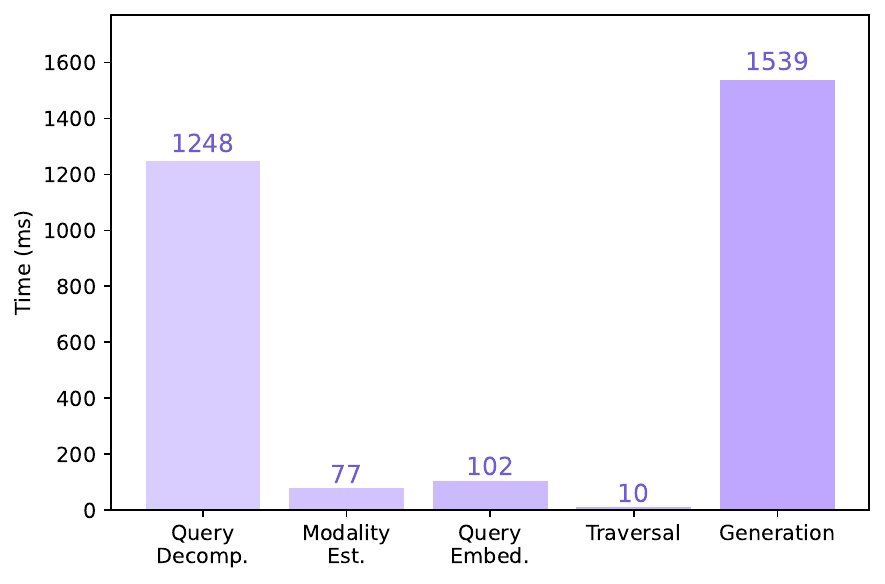}  \\
    (e) \texttt{InfoVQA} runtime comparison &
    (f) \texttt{LILaC}'s runtime breakdown on \texttt{InfoVQA} \\
    
    \runtime{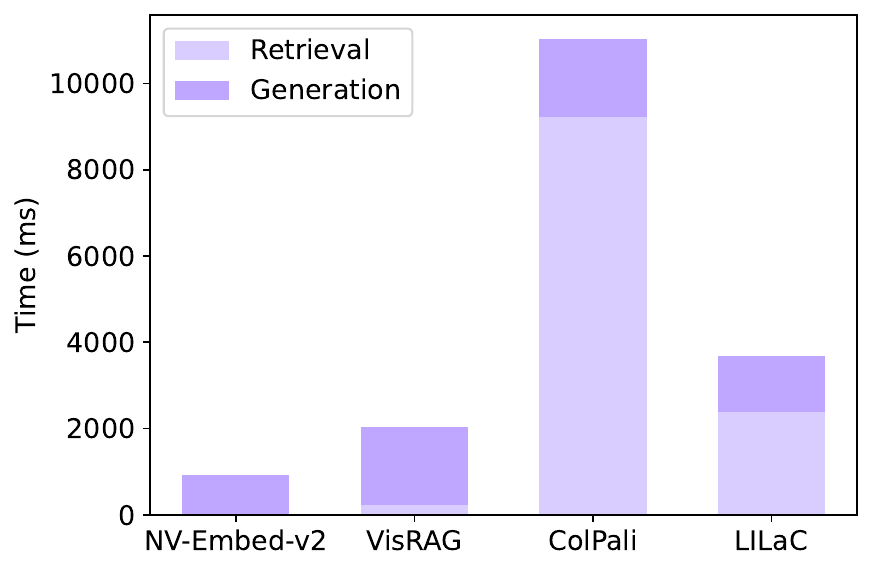}  & \breakdown{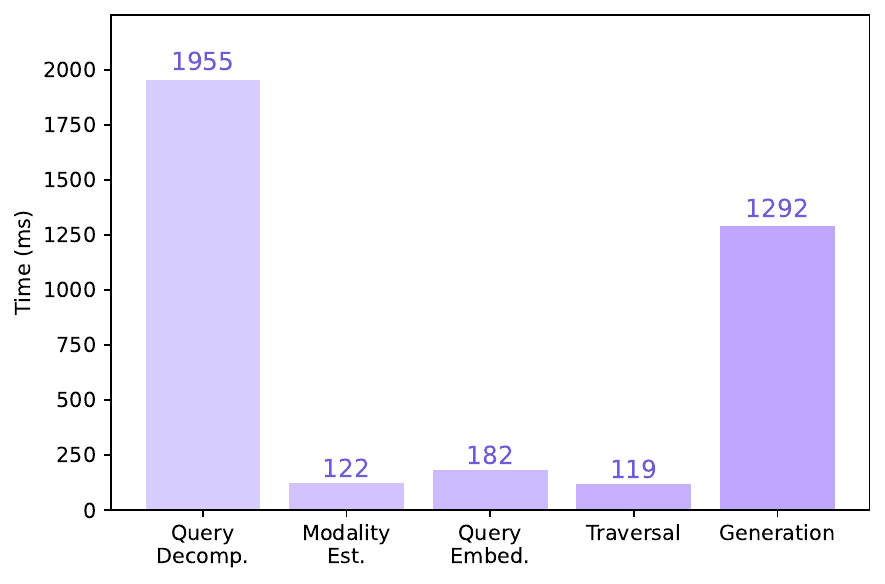}  \\
    (g) \texttt{MultimodalQA} runtime comparison &
    (h) \texttt{LILaC}'s runtime breakdown on \texttt{MultimodalQA} \\
    
    \runtime{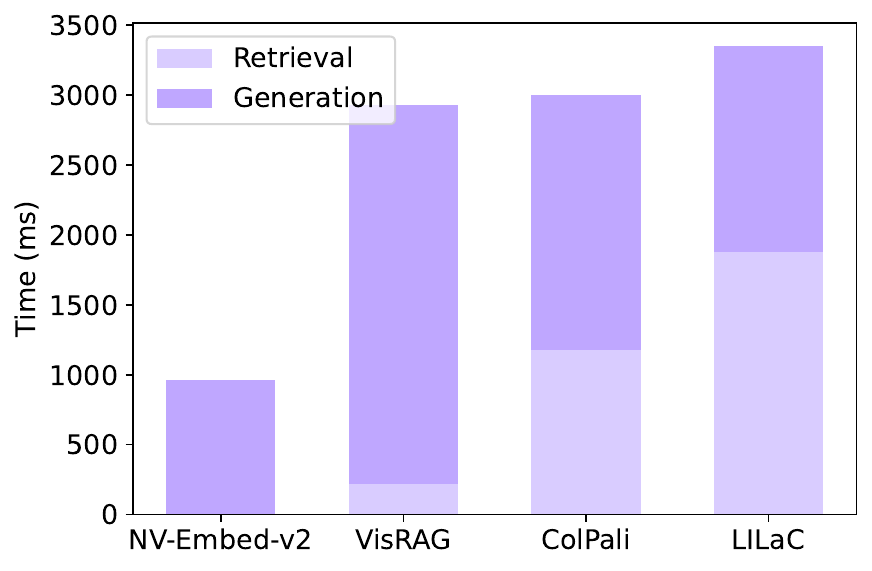}   & \breakdown{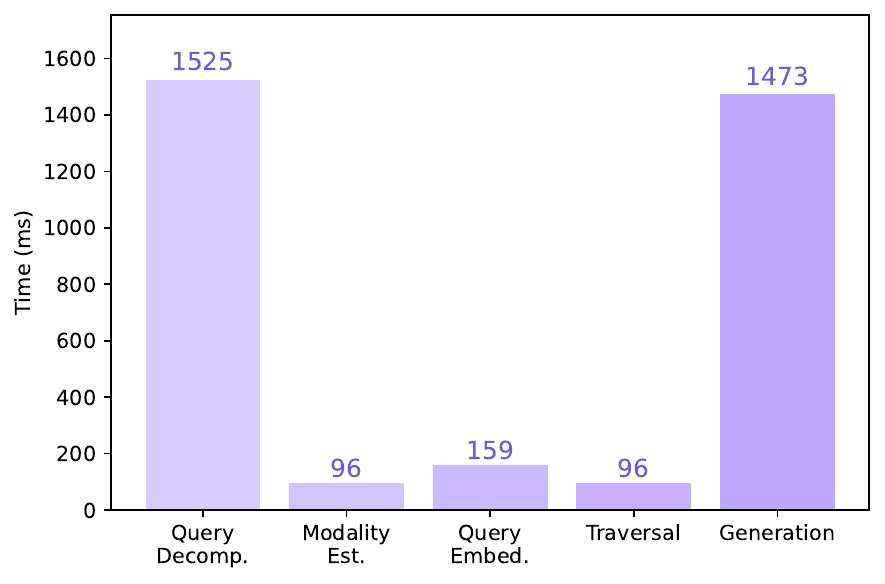}   \\
    (i) \texttt{MMCoQA} runtime comparison  & 
    (j) \texttt{LILaC}'s runtime breakdown on \texttt{MMCoQA} \\
    
  \end{tabular}
  \vspace{-3mm}
  \caption{Comparison of algorithm execution time (i.e., runtime) for each algorithm per dataset (left) and \texttt{LILaC}'s runtime breakdown per dataset (right).}
  \label{fig:appendix_runtime_breakdown}
\end{figure*}

We explored the impact of varying the beam width $b\ (\in \{1, 2, 3, 4, 5, 10, 20, 30\})$ on the retrieval accuracies.
As depicted in Figure~\ref{fig:parameter_sensitivity} (a), retrieval accuracy increased monotonically with larger beam widths, showing a significant improvement of 34.6\% in R@3 when expanding from the minimum of 1 to 30.
This trend highlights the benefit of wider beam searches, enabling more comprehensive and accurate graph traversal. 
Interestingly, despite these substantial accuracy gains, the overall execution time increased only marginally (2.8\%), indicating that graph traversal itself does not constitute the main computational bottleneck. 

Figure~\ref{fig:parameter_sensitivity} (b) presents retrieval accuracy as a function of iteration count $n_i$, varied from 0 to 2. 
We observed a modest yet meaningful 2.93\% improvement in R@3 when transitioning from zero to one iteration. 
This accuracy gain primarily results from enabling multihop reasoning, which is inherently unavailable at $n_i = 0$. 
While the overall increase might appear limited, it is particularly relevant to datasets explicitly requiring complex multihop reasoning, such as \texttt{MultimodalQA} and \texttt{MMCoQA}.

We further analyzed how varying key hyperparameters—beam width $b$ and the number of iterations $n_i$—affect the accuracy of \texttt{LILaC} across the five different datasets. 
We provide comprehensive plots illustrating the sensitivity and robustness of our method concerning these parameters in Figure~\ref{fig:appendix_param_sensitivity}.

\subsection{Layered Component Graph Construction Overhead}
\label{sec:lcg_construction_overhead}

\updated{
\textit{Theoretical complexity.}
We analyze the offline cost of building $\mathcal{G}$ (cf.\ Definition~\ref{def:layered_component_graph}). 
Let $n$ be the number of documents, $c$ the average number of \emph{components} per document, and $s$ the average number of \emph{subcomponents} per component.
The total cost decomposes as
\vspace{-1mm}
\begin{equation*}
\label{eq:tbuild-decomp}
    T_{\text{build}} = T_{\text{nodes}} + T_{\text{edges}} + T_{\text{embed}}
    \vspace{-3mm}
\end{equation*}
}

\updated{
\textit{Node generation.}
We enumerate components in each document and extract subcomponents for every component; we also add the containment links $(C,c)$ to $E_{\downarrow}$.
Enumerating all components across the corpus costs $O(nc)$, and extracting \& linking subcomponents costs $O(ncs)$:
\vspace{-1mm}
\begin{equation*}
\label{eq:tbuild-nodes}
    T_{\text{nodes}} = O(nc) + O(ncs)
    \vspace{-3mm}
\end{equation*}
}

\updated{
\textit{Edge generation.}
Within a document, we form the intra-document clique over $c$ components, yielding $\Theta(c^2)$ edges per document and $O(nc^2)$ overall.
Across documents, we follow the link mapping $\mathcal{L}$; with a hash map for document lookup, retrieving targets is $O(1)$ per link and contributes the same order.
Hence
\vspace{-1mm}
\begin{equation*}
\label{eq:tbuild-edges}
    T_{\text{edges}} = O(nc^2)
\end{equation*}
}

\updated{
\textit{Embedding generation.}
We embed all component and subcomponent nodes using $f$, which scales with their counts:
\begin{equation*}
\label{eq:tbuild-embed}
    T_{\text{embed}} = O(nc) + O(ncs)
    \vspace{-3mm}
\end{equation*}
}

\updated{
Summing the terms gives
\begin{equation*}
\label{eq:tbuild-final}
    T_{\text{build}} = O(ncs + nc^2)
\end{equation*}
In typical regimes where $c,s \ll n$, the offline construction is approximately linear in $n$.
The embedding term is usually dominant; importantly, it is fully offline, cacheable, and parallelizable across documents.
}

\updated{
\textbf{Empirical runtime.}
To validate the scalability, we measured average wall-clock time for each construction stage on increasing corpus fractions (20\%\,/\,40\%\,/\,60\%\,/\,80\%\,/\,100\%), holding the encoder $f$ and batching fixed.
Results are shown in Table~\ref{tab:lcg_offline_runtime}. 
They closely follow the above analysis, exhibiting near-linear growth in $n$ and revealing embedding as the primary bottleneck.
At 100\% of the data, total build time is 2h\,15m\,57s; embedding accounts for ${\sim}90.23\%$ of the cost, with node and edge generation contributing ${\sim}7.60\%$ and ${\sim}2.13\%$, respectively.
We emphasize that the offline cost can be further reduced via batching, sharding, and incremental updates when documents are added or modified.
}

\subsection{Comparison of Diverse Embedders}
\label{sec:embedder_comparison}

\updated{
To probe how biases in pretrained embeddings manifest in retrieval, we hold the \texttt{LILaC} pipeline fixed and vary only the multimodal embedder.
We evaluate seven families—\texttt{MM-Embed}, \texttt{UniME} (LLaVA-OneVision-7B-LoRA-Res336), \texttt{mmE5-mllama} (11B, instruct), \texttt{QQMM-embed}, \texttt{LLaVE} (0.5B/2B/7B), and \texttt{VLM2Vec} (Qwen2-VL; 2B/7B)—and report Recall@3 and MRR@10.}

\updated{
In Table~\ref{tab:embedder_bias}, we observe a consistent scaling trend: within \texttt{LLaVE}, Recall@3 improves by $+9.5\%$ relative from 0.5B to 2B ($62.01{-}56.59$ over $56.59$) and a further $+8.2\%$ from 2B to 7B; within \texttt{VLM2Vec}, 7B exceeds 2B by $+11.9\%$.
Overall, the top performers are \texttt{MM-Embed}, \texttt{UniME}, and \texttt{LLaVE}-7B.
These results indicate that LILaC's retrieval quality is sensitive to the inductive biases of the underlying encoder, yet benefits directly from stronger, larger models.
}

\subsection{Algorithm Execution Runtime: Further Analysis}

We conducted an in-depth examination of runtime efficiency. 
Specifically, we compared the overall execution time of our proposed method, \texttt{LILaC}, against other baseline algorithms across all datasets. 
We further broke down \texttt{LILaC}'s runtime into individual components (such as retrieval, reranking, and LLM refinement) to clearly identify performance bottlenecks and highlight the efficiency of different pipeline stages.
Detailed results are shown in Figure~\ref{fig:appendix_runtime_breakdown}.

\section{Prompt Templates}
\label{sec:prompt_templates}

We present detailed examples of the specific prompt templates used in our experiments. 
These prompts correspond to three key tasks: \textsc{Object Detection}, \textsc{Query Decomposition}, \textsc{Modality Selection} and \textsc{Answer Generation}. 
For each task, we provide clear instructions, expected input-output formats, and task-specific heuristics.

\begin{figure*}[htb]

\begin{tcolorbox}
[title = \textsc{Object Detection}, colback = gray!10, colframe = black, sharp corners, boxrule=0.5mm]

\textbf{Instruction}: \\
Detect all objects in the image and return ONLY a JSON list of {\texttt{\{class, bbox\_2d: [x1, y1, x2, y2]\}}}. Do NOT include markdown or extra text. \\
\\
\textbf{Image:} {\texttt{\{image\}}} \\
\textbf{Output:}
\end{tcolorbox}

\label{fig:object_detection}
\end{figure*}

\begin{figure*}[htb]

\begin{tcolorbox}
[title = \textsc{Query Decomposition}, colback = gray!10, colframe = black, sharp corners, boxrule=0.5mm]

\textbf{Instruction:} \\
You are a retrieval-oriented query decomposer. \\
\\
Goal – Produce the smallest set (1 – 5) of component-targeting sub-queries.  \\
Each sub-query must describe one retrievable component (sentence, paragraph, table row, figure, etc.) whose embedding should be matched.  \\
Together, the sub-queries must supply all the information needed to answer the original question. \\
\\
\textbf{Guidelines:} \\
1. Entity \& noun-phrase coverage: Every noun phrase and named entity that appears in the original question must appear at least once across the entire set of sub-queries (you may distribute them). 
Keep each phrase exactly as written. \\ 
2. One-component rule: A sub-query should reference only the facts expected to co-occur within the same component. If two facts will likely be in different components, put them in different sub-queries. \\ 
3. No unnecessary splitting: If the whole answer can be found in a single component, return only one sub-query. \\ 
4. De-contextualize: Rewrite pronouns and implicit references so every sub-query is understandable on its own. \\ 
5. Keyword distribution: Spread constraints logically (e.g., one sub-query for “light rail completion date”, another for “city with a large arched bridge from the 1997 Australia rugby-union test match”). \\ 
6. Remove redundancy: Merge duplicate or paraphrased sub-queries before you output. \\ 
7. Ordering for dependencies: If the answer to one sub-query is needed for another, place the prerequisite first. \\ 
8. Output format: Return only a JSON array of strings — no keys, explanations, or extra text. \\
\\
\textbf{Question:} {\texttt{\{question\}}} \\
\textbf{Output:}
\end{tcolorbox}
\label{fig:query_decomposition}
\end{figure*}

\begin{figure*}[htb]
\begin{tcolorbox}
[title = \textsc{Modality Selection}, colback = gray!10, colframe = black, sharp corners, boxrule=0.5mm]

\textbf{Instruction:} \\
You are a modality selector for multimodal QA. \\
\\
\textbf{Task:} \\
Given the single sub-question below, choose the one modality that is most appropriate for obtaining its answer. \\
\\
\textbf{Allowed modalities:}  \\
• text: unstructured prose (paragraphs, sentences, propositions) \\ 
• table: structured rows/columns (spreadsheets, stats tables, infoboxes) \\ 
• image: visual information (photos, posters, logos, charts) \\
\\
\textbf{Heuristics:}  \\
1. Numeric totals, percentages, year-by-year figures $\rightarrow$ table \\ 
2. Visual appearance, colours, logos, “what does … look like” $\rightarrow$ image \\ 
3. Definitions, roles, biographies, causal explanations, quotes $\rightarrow$ text \\ 
4. If two modalities could work, pick the one that will yield the answer fastest. \\
\\
\textbf{Output format:} \\
Return only the modality label on a single line – exactly \texttt{text}, \texttt{table}, or \texttt{image}. \\ 
No JSON, no additional text. \\
\\
\textbf{Subquery:} {\texttt{\{subquery\}}} \\
\textbf{Output:}
\end{tcolorbox}
\label{fig:modality_selection}
\end{figure*}

\begin{figure*}[htb]
\begin{tcolorbox}[
  enhanced,
  title = \textsc{Answer Generation},
  colback = gray!10,
  colframe = black,
  sharp corners,
  boxrule = 0.5mm
]

\textbf{Instruction:}\\
Using the \texttt{f\_answers()} API, return a list of answers to the question based on \emph{retrieved webpage components}.
A retrieved component can be a passage, a table, or an image.
Strictly follow the format of the example below and keep the answer short.
For \emph{yes/no} questions, respond only with \texttt{f\_answers(["yes"])} or \texttt{f\_answers(["no"])}.\\

\textbf{Example:}\\[-2pt]
\noindent\rule{\linewidth}{0.5pt} 

\texttt{[Passage]}\\
Document title: South Asia\\
The current territories of Afghanistan, Bangladesh, Bhutan, Maldives, Nepal, India, Pakistan, and Sri Lanka form South Asia. The South Asian Association for Regional Cooperation (SAARC) is an economic cooperation organisation established in 1985 that includes all eight nations comprising South Asia. \\

\texttt{[Passage]}\\
Document title: UK Joint Expeditionary Force\\
The UK Joint Expeditionary Force (JEF) is a United Kingdom-led expeditionary force which may include Denmark, Finland, Estonia, Latvia, Lithuania, the Netherlands, Sweden, and Norway. It is distinct from the Franco-British Combined Joint Expeditionary Force. \\

\texttt{[Table]}\\
Document title: Lithuanian Armed Forces — Current operations\\[2pt]
\TextHeader
\TextRow{Somalia}{EU}{Operation Atalanta}{15}
\TextRow{Mali}{EU}{EUTM Mali}{2}
\TextRow{Afghanistan}{NATO}{Operation Resolute Support}{29}
\TextRow{Libya}{EU}{EU Navfor Med}{3}
\TextRow{Mali}{UN}{MINUSMA}{39}
\TextRow{Iraq}{CJTF}{Operation Inherent Resolve}{6}
\TextRow{Central African Republic}{EU}{EUFOR RCA}{1}
\TextRow{Kosovo}{NATO}{KFOR}{1}
\TextRow{Ukraine}{—}{Training mission}{40}

\vspace{4mm}
\textbf{Question:} Among the Lithuanian Armed Forces' current operations, which deployment involves fewer personnel: Kosovo, or the deployment in the nation that, along with six others, constitutes the sub-continent of South Asia? \\
\textbf{Answer:} The South Asia passage shows Afghanistan is part of that region. The table lists 29 personnel in Afghanistan and only 1 in Kosovo, so \texttt{f\_answers(["Kosovo"])}.
\noindent\rule{\linewidth}{0.5pt} 
\\ \\
Using the images and texts given, answer the question below in a single word or phrase.\\ \\
\texttt{\{retrieved components\}}\\ \\
\textbf{Question:} \texttt{\{question\}}\\
\textbf{Answer:}
\end{tcolorbox}
\label{fig:answer_generation_prompt}
\end{figure*}

\end{document}